# Toroidal Plasmonic Meta-Biosensors: Benchmarking against Classical Plasmonic Biosensors and Transducers

Arash Ahmadivand,[1,*] Burak Gerislioglu,[1]

[1]Department of Physics and Astronomy, Rice University, 6100 Main St, Houston, Texas 77005, United States

*aahmadiv@rice.edu

**ABSTRACT:** Rapid progress in plasmonics biosensors and immunosensors brings many methodologies to employ resonant subwavelength structures associated with the physics of multipoles, Fano resonances, and analogous moments. Here, we review a broad range of traditional and novel techniques that have been utilized for developing advanced biosensing tools including colloidal nanoparticles (NPs)-based systems to well-engineered arrays of meta-atoms and metamolecules. By describing some of the important and practical label-free biosensing methods based on subwavelength plasmonic technology, we discuss the emerging and excitation of toroidal multipoles concept, as a third family of multipolar modes, has made a revolution in enhancing the sensing performance of ultra-low weight diseases proteins and viruses as biological targets. Taking the exquisite and untraditional advantages of toroidal plasmonic metasurfaces, we comprehensively evaluated the detection properties and high-quality of toroidal immunosensors for the sensing of signature of bioproteins at picomolar (pM) concentrations with excellent limit of detection (LOD). This understanding clearly shows how toroidal metasensors have emerged as promising clinical and pharmacology tools, can be considered as alternatives and big competitors for conventional plasmonic sensors and transducers.

**KEYWORDS:** plasmonics, biosensing, metamaterials, Fano-resonant sensors, toroidal resonances, toroidal immunosensors.

# 1. Introduction

Surface plasmon resonances (SPRs) are coherent *d*-band electron oscillations, occurring at metal-dielectric interfaces, when exposed with intense light of certain frequencies.[1,2] As a promising counterpart of optical physics and nanophotonics for extreme light confinement and manipulation,[3] plasmonics has been acknowledged as an inspiring notion for developing advanced and practical photonics technologies, devices and applications.[4-6] Possessing fundamental role for designing several optics-based and optoelectronic devices, plasmonics has enabled designing of efficient light harvesters and photovoltaic devices,[7-10] long-decay range hybrid waveguides,[4,11,12] ultrafast modulators,[13-15] photodetectors and transistors,[16-20] polarization beam splitters,[21-23] Mach-Zehnder interferometers,[24-26] metamaterials,[27-32] superlensing,[33,34] nanolasers,[35-37] quantum devices,[38-40] solar water splitting,[41,42] bio-related devices and clinical tools,[43-45] etc. For the latter case, promisingly, plasmonics is a reliable concept with exquisite spectral features, in both *in vitro* and *in vivo* assays, has been utilized for developing several types of label-free diseases and viruses diagnosing devices,[46-55] targeted cancer and tumor therapies,[56-60] directed drug delivery,[61-64] nanowelding,[65,66] hyperspectral nano-imaging,[67-69] optoacoustic imaging,[70,71] real-time pharmacology,[72-74] vapor and micro-bubble generation,[75-77] laser nanosurgery,[78-80] photothermal heat spectroscopy,[81-85] photothermally controlled fluidics,[86-89] heat-assisted magnetic recording,[90-93] and optical monitoring and detection of DNA.[94,95] Extreme localization of SPRs in metallic nano-objects leads to robust enhancement in the optical absorption, accompanied by thermal heating of the free electron gas *via* electron-electron scattering in a hundreds of femtoseconds.[96] Taking the advantage of substantial localization and confinement of plasmons in metallic subwavelength particles, and also significant absorption cross-section, new doors have opened to tailor various useful biodevices.

Of particular interests are the plasmonic biological and biochemical sensors, which possess key role in commercial and advanced clinical and pharmaceutical applications.[43,44,97,98] High accuracy, real-time response, label-free, operating in room-temperature, cost-effective, and fast response, all triggered researchers to work on enhanced plasmonic biosensors for decades.[99,100] In addition, plasmonic nanodevices provide non-destructive and non-poisonous interaction with biological tissues in the corresponding assays.

All these features enabled plasmonic integrated sensors to be a strong rival for the traditional sensing methods in medical applications such as the enzyme-linked immunosorbent assay (ELISA),[101,102] point of care testing (POCT),[103] electrochemical sensing,[104-106] and reverse transcription polymerase chain reaction method (RT-PCR).[107] Although these sensing tools are the most common approaches in daily life, but plasmonics and other optical sensing technologies have emerged as potential alternatives for these methods, as the next-generation and advanced precise infection and virus detection.[43,108]

Technically, most of the plasmonic biochemical and biological sensors and transducers are operating based on the the spectral response of the optically induced resonant modes. To begin with, noble metallic (Silver (Ag), Gold (Au), Copper (Cu), etc.)[109] nanoparticles (NPs) in different shapes (i.e. spheres,[110] rods,[111] rings,[112] matryoshkas,[113] starts,[114] etc.) with distinct dipolar modes have been employed for biological sensing purpose due to having strong binding properties of biomolecules to the surface area of the NPs. However, despite bearing promise for novel technological advances, these NPs are quite far from being highly sensitive to the minor environmental perturbations and low concentration of proteins, with regards in particular to the origin and nature of their broad plasmonic resonances. Seeking for resonant-structures which hold highly narrow and sharp spectral lineshapes (i.e. Fano resonances and electromagnetically induced transparency (EIT)), has led researchers to design artificial and well-engineered structures such as NPs assemblies,[115-120] and meta-atoms.[121-127] Both Fano- and EIT-resonant nanostructures have attracted copious interests for developing compact and portable biological and clinical immunosensors.[128-130] Although current plasmonic sensors are promising and provide substantial sensitivity and reasonable limit of detection (LOD), these technologies are still quite faraway from very early stage diagnosis of ultra-low weight infections and viruses at slight concentrations. Early detection of specific illnesses such as, Alzheimer's and Parkinson's diseases, and pestiferous disorders is extremely critical for the avoidance of huge life and medical expenses.[131]

Very recently, an alternative approach has been proposed and experimentally validated by Ahmadivand and colleagues for the detection of biological targets at ultra-low concentrations (in the range of a few pg/mL) based on plasmonic metasensors.[132,133] Such extremely sensitive metadevices have been developed based on a new family of multipolar resonant moments, known as toroidal multipoles.[134-136] Theoretically, static toroidal dipole was introduced for the first time in 1957, in the context of nuclear physics.[137] Instead, dynamic toroidal dipole, driven by intense incident beam illumination, is an independent term and third member in the family of electrodynamic multipole expansion. Possessing analogous properties to the dark-side of plasmons,[138] toroidal multipoles, as a rising trend, have received growing interest for several practical applications because of facilitating negligible radiative losses due to suppression of electric dipole.[139] Among all these exquisite advantages, the hybrid behavior, narrow lineshape, and ultrahigh sensitivity of toroidal dipole-resonant metamaterials to the environmental perturbations have led to the emerging of several optical and biological-sensing tools with unique spectral properties.[140-142] The physical mechanism behind and sensing properties of these particular resonances will be discussed in the forthcoming sections.

## 2. Classical electromagnetic multipolar-resonant colloidal NPs for biosensing

Noble metal NPs in different shapes and colloidal arrangements are traditional choices for developing attitudes for the detection of various diseases, treating tumors, and neuron stimulation, ranging across the visible to the near-infrared region (NIR).[56,74,143-145] Under intense beam illumination and at the resonant wavelength, strongly localized surface plasmon resonances (LSPRs) forming around and between the proximal NPs aggregates, show strong dependency to the environmental media dielectric variations.[146] Dipolar LSPR (DLSPR) is the fundamental member of classical electromagnetic multipoles family, can be observed and identified as a distinct peak in either extinction and scattering cross-sectional profiles for NP aggregates. Any variation in the refractive index (RI) of the surrounding media of the colloidal NPs agglomeration red-shifts the position of the DLSPR to the lower energies (longer wavelengths). It is well-accepted that the position of DLSPR depends on the shape, geometry, and material of NPs aggregate.[146,147] In this regime, the amount of resonance shift due to RI variations determines the sensitivity of the system. In the following subsections, we will separately investigate the plasmonic properties and sensing capabilities of spherical and nonspherical NPs.

## 2.1. Spherical NPs

In this section, we firstly analyze the unique properties and advantages of spherical shape NPs. Possessing fundamental role in understanding the concept of plasmonics and optical systems, these type of NPs have broadly been employed for both biochemical and biological sensing purposes. Mock and co-workers,[146] as a pioneer work, have experimentally verified the dependency of the plasmonic resonant moments excited from Ag colloids to the local RI variations of the ambience. In terms of optical physics, scattering theory for NPs[148] is the approach to predict the spectral response of spherical objects. Using this method and employing finite difference time domain (FDTD) analysis,[149] we qualitatively computed and plotted the scattering cross-section for spherical Ag NPs with varying diameters from 40 nm to 90 nm numerically, shown in Figures 1a and 1b. Following the experimental analyses in Mock's work, for the NP in the air atmosphere with RI of $n=1$, a distinct DLSPR arises around ~345 nm for the colloids with the diameters of 40 nm. Increasing the diameter of the NP to 90 nm, gives rise to slight red-shifts in the position of dipole and enhances the intensity of mode (Figure 1a). By switching the material around the colloidal NPs from air to oil with the index of $n=1.44$, a global red-shift happens in the position of dipolar modes in all geometries (Figure 1b). Obviously, all peaks are red-shifted to the lower energies ranging between 375 nm to 395 nm, for the diameter of 40 nm to 90 nm, respectively. Technically, the electrostatic model can be applied to understand the origin of the DLSPR mode shift ($P_{DLSPR}$):[150]

$$P_{DLSPR} \sim \frac{\varepsilon(\lambda) - \varepsilon_{media}}{\varepsilon_{media} - d_p \left( \varepsilon(\lambda) - \varepsilon_{media} \right)} \tag{1}$$

where $d_p$ is the depolarization factor (which is 1/3 for spherical NPs), and $\varepsilon_{media}$ is the permittivity of the surrounding media. Therefore, in the resonant condition, $\varepsilon(\lambda)$ expresses as $\varepsilon(\lambda) = \left(1 - 1/d\right)\varepsilon_{media}$, in which differentiating this equation, using the depolarization factor for a sphere, and the relation between RI and permittivity ($n_{media}^2 = \varepsilon_{media}$), one can write:

$$\frac{d\varepsilon(\lambda)}{d\lambda_r} = -2 \frac{d\varepsilon_{media}}{d\lambda_r} \tag{2}$$

Considering the empirically measured permittivity values for Ag as a function of wavelength by Palik[151] and Johnson-Christy,[152] one can define the derivative on the left-side of the equation above. Thus a change in resonant frequency with index shift of roughly 0.7 nm per change in the index of 0.01, near a wavelength of 500 nm is expectable.

In the experimental assays, Mock and co-authors have observed analogous trend in their analyses. Figures 1c and 1d represent dark-field images of Ag NPs with the average diameter of ~70 nm (ranging between 40 nm to 90 nm), immobilized on $SiO_2$ wafer with 100× objective before and after immersing the specimens in oil. It should be noted that radiation of different colors from the system is due to the excitation of variety of plasmonic modes in the visible spectrum. This implies that the particles with blueish hue are perfectly spherical, and the others are triangle or other nonspherical shapes. In this set of experiments and measurements, the Ag NPs were immobilized on polylysinetreated silicon wafers (having ~100 nm $SiO_2$ top layer to insulates the plasmonic NPs from the conductive silicon (Si) substrate), at low density, by placing a drop (50-500ul) of the colloidal solution onto the wafer for ~1 min. The polylysine treatment firmly immobilizes the colloidal Ag NPs to the substrate surface. Figure 1e exhibits the spectral response and DLSPR shift as a function of wavelength for roughly spherical NPs with the average diameter of 70 nm. It should be noted that there is a few nanometers difference between the shifted dipole in experiments and simulation, which caused by either experiment imperfections or carrying out 2D simulations by us.

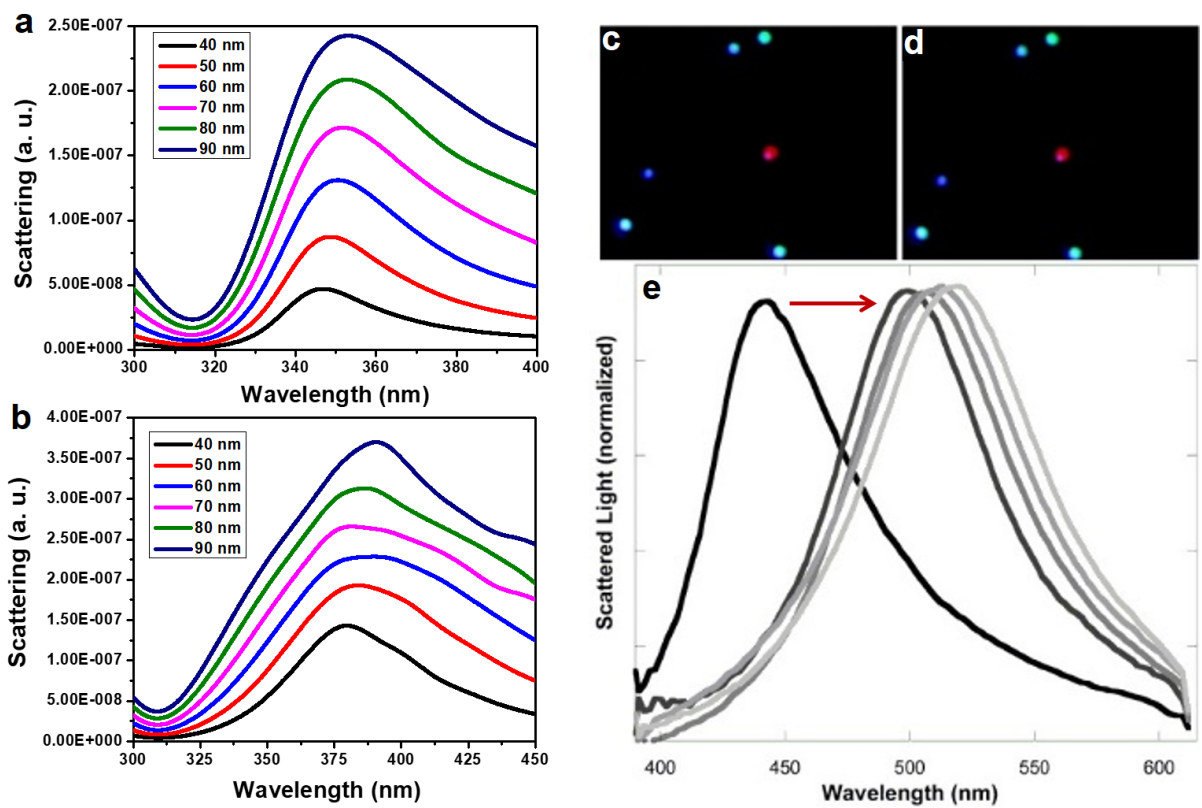

**Figure 1.** Numerically obtained spectral response and scattering cross-section of Ag nanospheres in a) air, and b) oil media, respectively. c, d) Typical field of Ag NPs immobilized on $SiO_2$ wafer and imaged under dark-field illumination with 100× objective. e) Experimentally measured spectral shift for individual blueish spherical Ag NPs. Typical blue particle spectrum as it is shifted from air to 1.44 RI oil, and successive oil treatments in 0.04 RI incremental increases. The arrow in the panel show the shift of the resonance from air ambience to oil media.[146] Copyright 2003, American Chemical Society.

Having reasonable shift in the dipole moment position facilitates huge surface area for strong binding of biological objects (i.e. proteins, and antibodies), easy synthesis, and quick simulation, enabled nanospheres to have extensively for several bio-related applications.[153,154] In spite of providing such benefits, these attributes are not enough for developing advanced clinical and precise biochemical and biomolecules detection tools and solutions. Among them, broad DLSPR lineshape and weak hotspots dramatically reducing the detection performance and accuracy of the corresponding systems. To address these limitations, nonspherical NPs with the ability to support edge plasmons,[155,156] and much stronger localized plasmons at the outermost tips and margins have been investigated, and still scientists utilizing these objects in their real-time analysis and assays.

**2.2.Nonspherical NPs**

The field enhancement in plasmonic nanoplatforms plays fundamental role in developing highly sensitive and efficient environmental sensing tools. Both the LSPR and charge transfer between the molecule and the metal *d*-band are the origins for such a feature.[157] As it was mentioned and explained in the previous section, the enhancement for a single isolated Ag NP is less than four orders of magnitude and the intensity of LSPR moment is not strong enough for advanced biosensing an transducing purposes. Further enhancement in the operating quality of label-free nanosensors can be obtained by inducing robust hotspots, which are highly localized surface plasmons, that are producible from sharp protrusions (i.e. rods, stars, cones, etc.),[145,155,156,158] or coupled nano-configurations (e.g. dimers, trimers, aggregates, etc.).[159-161] Nonspherical nanostructures allows for tailoring the corresponding geometry, enable possessing tunable, and selective excitation of strong hotspots to optimize the spectral response. Among all of visible and NIR-tuned nonspherical NPs, metallic nanostars are strategic and promising NP colloids for bioapplications, have been utilized extensively for enhancing the Raman signal up to ~$10^{14}$ in surface–enhanced Raman spectroscopy (SERS) spectral characterization,[159,162] and selective generation of hotspots.[114,161,163] Nanostars provide highly tunable plasmon bands along the wide range of spectrum including multiple narrow lineshapes acting as hotspots.[164] Furthermore, nonspherical NPs inherently acting analogous to the antisymmetric structures and reflect strong polarization-dependency. Following the same trend, the LSPR peak position and the field confinement efficiency in nanostars are associated with the branch aspect ratio and branch length-number, respectively.

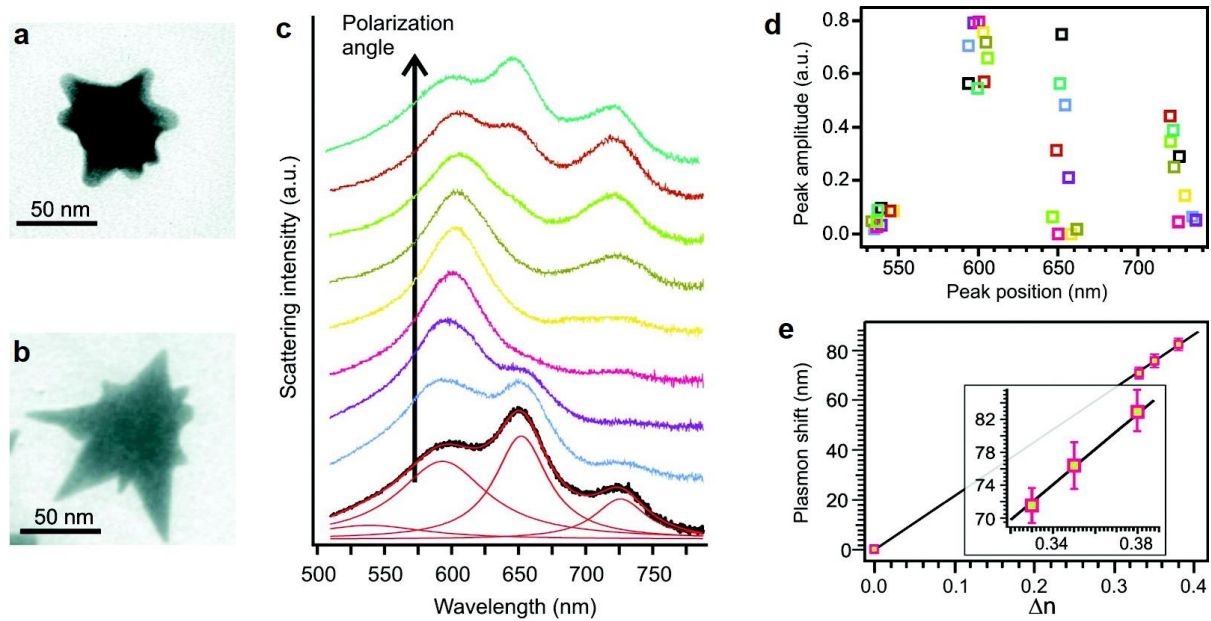

**Figure 2.** a and b) Transmission electron microscopy (TEM) images of two gold nanostars. c) Polarized light scattering spectra of a single gold nanostar immobilized on a glass substrate (in 0.1 M PBS buffer solution; pH 6.8) as a function of the polarization angle (0°-180°) of the incident light. A four-peak (Lorentzian) fit to the first spectrum is shown. d) Scatter plot of the amplitude vs position obtained from four-Lorentzian fits to the spectra shown in panel c. e) Average spectral shift of nanostars resonances peaking between 650 and 750 nm as a function of the surrounding refractive index. Data points for air ($n$=1.00), Millipore water ($n$=1.33), and 0.35 and 0.95 M glucose water solutions ($n$=1.35 and $n$=1.38, respectively).[167] Copyright 2010, American Chemical Society.

As a leading study in using plasmonic nanostars for label-free and single particle biosensing,[167] Dondapati and colleagues have shown that a single nanostar can be tailored to support multiple plasmon resonances corresponding to the tips and core-tip interactions. To this end, the authors have carried out experimental assays to detect Streptavidin (SA) molecules *via* strong binding with biotin-modified Au nanostars by spectral shifts in the plasmon resonances. Figures 2a and 2b illustrate transmission electron microscopy (TEM) images for the fabricated plasmonic nanostar. Sustaining multiple LSPRs (plotted in Figure 2c), substantial polarization dependency is monitored in the scattering intensity profile. This cross-section demonstrates the excitation of multiple Lorentzian resonances along the visible band. Figure 2d represents a scattering diagram for the amplitude as a function of the optically driven peak position obtained from four-Lorentzian fits to the polarization dependent spectra of Figure 2c and is the characteristic of all nanostars investigated. In this regime, one can observe the excitation of a weak and almost polarization-independent resonance around 540 nm-560 nm, supporting by the core of each nanostar. The polarization-dependency of the excited resonances at lower energies (longer wavelengths) associating with the outermost tips or the interaction between the tips and cores substantially enhances. This is because of having direct interference between the localized fields and surrounding media.

To analyze the spectral shift of the optically excited LSPRs during bulk variations in the RI of the medium, the corresponding spectra of individual nanostars in air, water, and glucose solutions of different concentrations are measured (Figure 2d). By quantifying the shift of each resonance extreme with a four-peak (Lorentzian) fit to each spectrum, one can estimate the sensitivity of the system to the environmental perturbations. To this end, an average response for a given nanostar is included in the computations, from different nanostars peaking between 650 nm and 750 nm. Relatively, the obtained average sensitivity is 218 nm/RIU, shown in Figure 2d. Considering the average full width at half maximum (FWHM) of 43 nm computed for the resonances, thus the corresponding figure of merit (FOM) can be assessed around ~5.

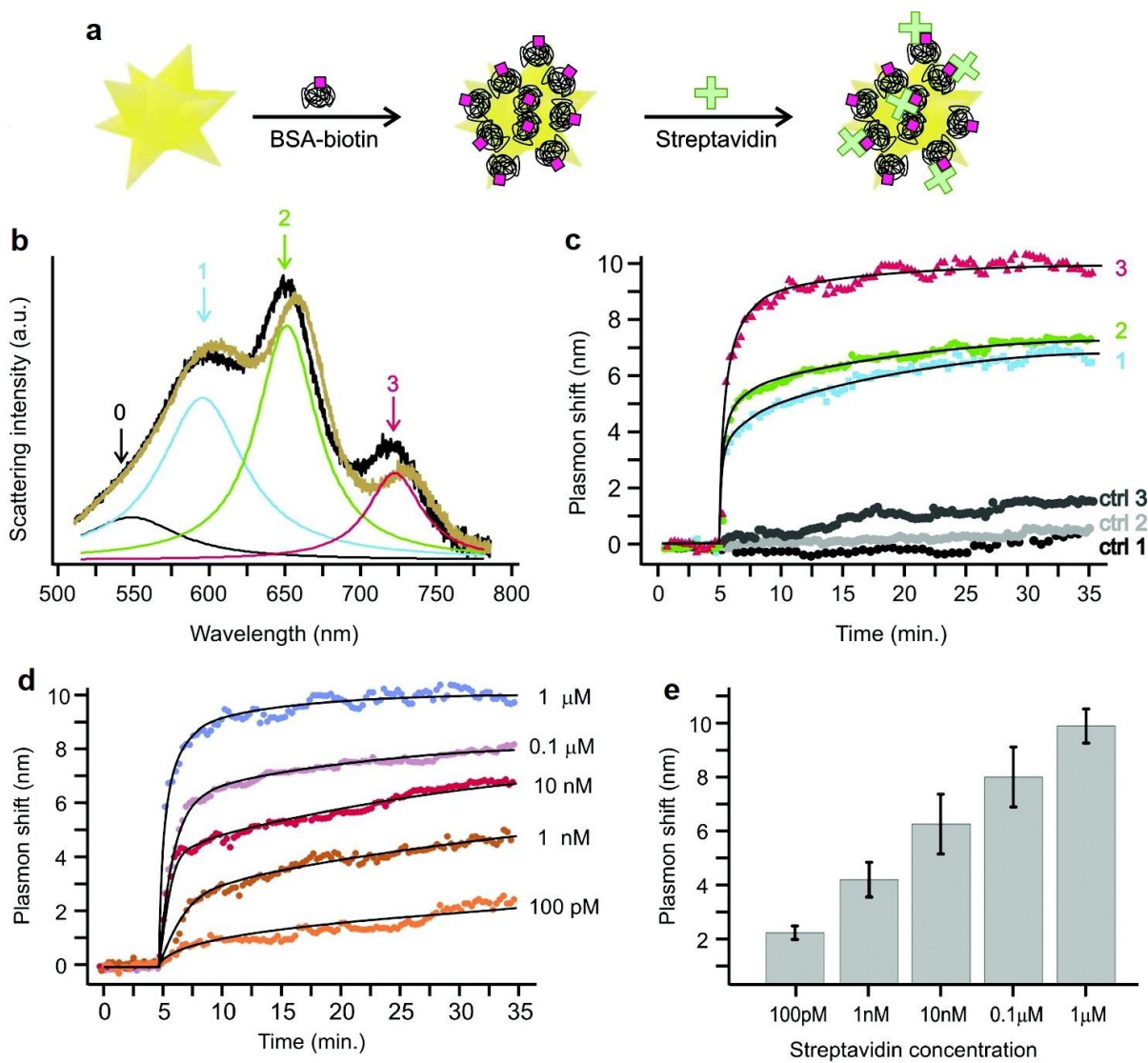

**Figure 3.** a) Schematic representation of SA-biotin interactions on a single Au nanostar. b) Scattering spectra of a single gold nanostar before (black) and 30 min after the incubation in a 1 m SA solution (brown). The curves are the Lorentzian peaks obtained from the fit to the spectra before incubation with SA. c) Time-dependent plasmonic extreme shifts corresponding to the three peaks of the biotin-modified Au nanostar shown in (**b**) for 1 m SA concentration. Typical control curves obtained for nanostars resonances peaking between 650 and 750 nm: (ctrl 1) pure buffer, (ctrl 2) addition of 1 m BSA, (ctrl 3) response of an Au nanostar modified with nonbiotinylated BSA to a 1 m SA solution. d) Real time measurements of the plasmon peak shift of the LSPR "3" of single biotin-modified Au nanostar for different SA concentrations. e) Average shifts as a function of the SA concentration (five different nanostars for each concentration).[167] Copyright 2010, American Chemical Society.

Following, we discuss the practical application of the plasmonic nanostar platform for the detection of SA through establishing binding sites over the surface area and branches of biotinylated Au nanostars (Figure 3a). Using biotin-functionalized nanostars and binding the SA biomolecules to the biotin moieties on the NP surface, the local RI of the surrounding media increases, giving rising to significant red-shift in the position of the plasmonic moments to the longer wavelengths. Here, Dondapati and colleagues quantified the plasmon shifts with a four-peak (Lorentzian) proper to the spectra of the single Au nanostars (Figure 3b). Relatively, the peak labelled as "0" attributed to the plasmon of the nanostar's core, is polarization independent and reflects the most trivial spectral response to the presence of SA molecules. Focusing over the more sensitive peak, they calculated the shift in the wavelength units and perceived an explicit trend. The longer wavelength peaks respond to the presence of SA target molecules with stronger shift (Figure 3c), consistent with the previous reports and theoretical calculations in the literature.[168,169] Having largest detectable wavelength shift is a valuable advantage in a given biosensor, and in the current nanostar, the spectral shift of the excited peaks 1, 2, and 3 is 1.10%, 1.11%, and 1.36%, respectively. This trend strongly verifies the highest sensitivity of the lower energy peak around ~725 nm to the dielectric variations of the ambient. By carrying out a series of typical controlled experiments, one can control the stability of the LSPRs in the buffer solution. Accordingly, no shifts are observed in ctrl 1. Then, the nanostars' response to 1 m BSA solution is tested. BSA has a molecular weight similar to SA and hence the SA and BSA solutions have comparable RIs. Only a trivial shift is observed at long incubation times due to unspecific binding (ctrl 2). Ultimately, an extra control measurement has been performed with Au nanostars modified with nonbiotinylated BSA. This results with an LSPR shift of around ~1.5 nm is detected due to the nonspecific adsorption of SA (ctrl 3).

Finally, let us focus on the sensitivity of an Au nanostar to the presence and binding of different concentrations of SA biomolecules. The real-time monitoring of the shift of plasmon resonances of single biotinylated nanostars incubated in different concentrations of SA biomolecule targets is plotted in Figure 3d. Including the right-most resonance peak (lower energy peak) in the analysis, by pipetting of 1 m SA to the system after 5 min, there is a severe red shift of the plasmon peak within 2 min, reaching a saturation limit of about 10 nm. Lower concentrations of SA biomolecules require more time to yield smaller shifts: for 0.1 m, 10 nM, and 1 nM, there is a steady-state plasmon resonance shift of around 8, 6.5, and 4.2 nm. For the lower concentrations of targeted bio-agents, a clear saturation was not observed even after 30 min. The average plasmonic peaks shifts taken from five different nanostars as a function of the SA are explicitly demonstrates in Figure 3e. The Au nanostars are highly sensitive to the presence of analyte leading to a very limited linear-response range (more than 1 nM). The lowest concentration or in other words the LOD of the detection method was estimated around 0.1 nM, has been observed for plasmon resonance shift around 2.3 nm.

Although the proposed approach based on metallic nanostars is promising, but it suffers for ultra-low weight biomolecules recognition at very small concentrations. For instance, in the study above, the corresponding molecular weight of SA protein is around 52.8 kDa, which is detected for the concentration of a few nM. However for the practical clinical purposes and early-detection of diseases, having much more sensitive nanosystems to the presence of ultra-low concentration of bio-objects is preferred. This shortcoming has been successfully addressed by developing well-engineered nano and microstructures that are able to support sharp and narrow plasmonic resonances with large shifts in the position of the induced moments to the environmental perturbations.

## 3. Fano-resonant sensors

Living in the world of resonances and driven by the need, of particular interest are ultrasensitive plasmonic biosensors based on antisymmetric resonance lineshapes. In the past decade, there have been great efforts to develop precise and fast plasmonic nanosensors based on "the dark side of plasmonics".[138,142,170] Introduced for the first time by Prodan and colleagues,[171] the dark side of plasmonics or the antibonding modes have been observed in intensely coupled and hybridized plasmonic nanosystems. Such an effect has been successfully experienced in simple and fundamental NP clusters and artificial aggregates such as two- and three-member dimers and trimers, respectively.[172,173] These ensembles of NPs enabled efficient and tunable manipulation and confinement of electromagnetic fields at the subwavelength level.[170-173] Breaking the symmetry of plasmonic nanostructures or increasing the complexity of artificial nanoassemblies lead to support traditional electric and magnetic multipoles as well as additional antisymmetric lineshapes such as Fano-like and Fano resonances.[115-122,174] The direct interference between the dark (subradiant) and bright (superradiant) modes gives rise to the formation of Fano resonances, characterized by a narrow spectral transparency window where scattering is substantially suppressed and absorption is boosted.[174]

### 3.1. The theory of the dark-side of plasmons

Plasmonic dark modes (Subradiant modes) are pure near-field moments that can arise from the closely-packed and hybridized NPs. In comparison to bright superradiant modes, dark modes possess longer dephasing time because of the absence of electric dipole moment, which making them loss-less and efficient for various nanophotonics applications. Conversely, bright superradiant modes dramatically suffer from radiative losses, which reducing the efficiency of the developed devices. Theoretically, the strong confinement of optical energy in these modes and their accompanying near-fields hold great promise for achieving strong coupling to single photon emitters.[138] In the nonretarded limit (quasistatic approximation) and in the hybridized nanosystems, dark resonances are the modes arising from high-order multipolar resonances supported by the NP-based system (i.e. the quadrupolar modes that exist in). In this limit, the dark modes do not easily couple to the incident beam. On the other hand, dark modes also arise from the interaction of the bright modes of coupled NPs. As a fundamental and simple member of NP clusters family, a two-member dimer consisting of a pair of NPs, can be tailored to support a dark mode originate from an out-of-phase oscillation of the bright (dipolar) modes of the particles.[175-177] As we mentioned earlier in this section of the current article, this results in the formation of a collective plasmon mode without a dipole moment.

A simple example allows us to understand the physical mechanism behind the formation and behavior of subradiant dark modes in the hybridized plasmonic systems. Figures 4a and 4b show schematic diagrams for the plasmonic hybridization of a nanodimer system composed of a pair of nanoshells positioned in distant and closely-coupled regimes, developed by Lassiter and colleagues.[178] Under *x*-polarized beam illumination, the effective hybridization of two levels is determined by the ratio of the square of their interaction energy and their energy difference. For the dimer in weak coupling phase and distant NPs (with the gap of ~20 nm), the interaction between the proximal nanoshells results in a trivial splitting of the nanoshell modes into bright superradiant and dark subradiant levels, which essentially retain the same multipolar index (*l*). In this regime, only the lowest order dipolar *l*=1 mode (mode 1) is excitable by the incident beam because this mode has a strong dipole moment merely (Figure 4a). Reducing the interparticle distance between nanoshells increases the coupling intensity and the splitting into bright and dark modes becomes much larger than for the weak interaction case (Figure 4b). Consequently, despite of the difference in energy between nanoshell plasmons of different multipolar *l*, the intense interaction results in the hybridization of different multipolar order plasmons such that each dimer mode contains an admixture of different *l* nanoshell modes.[175,176]

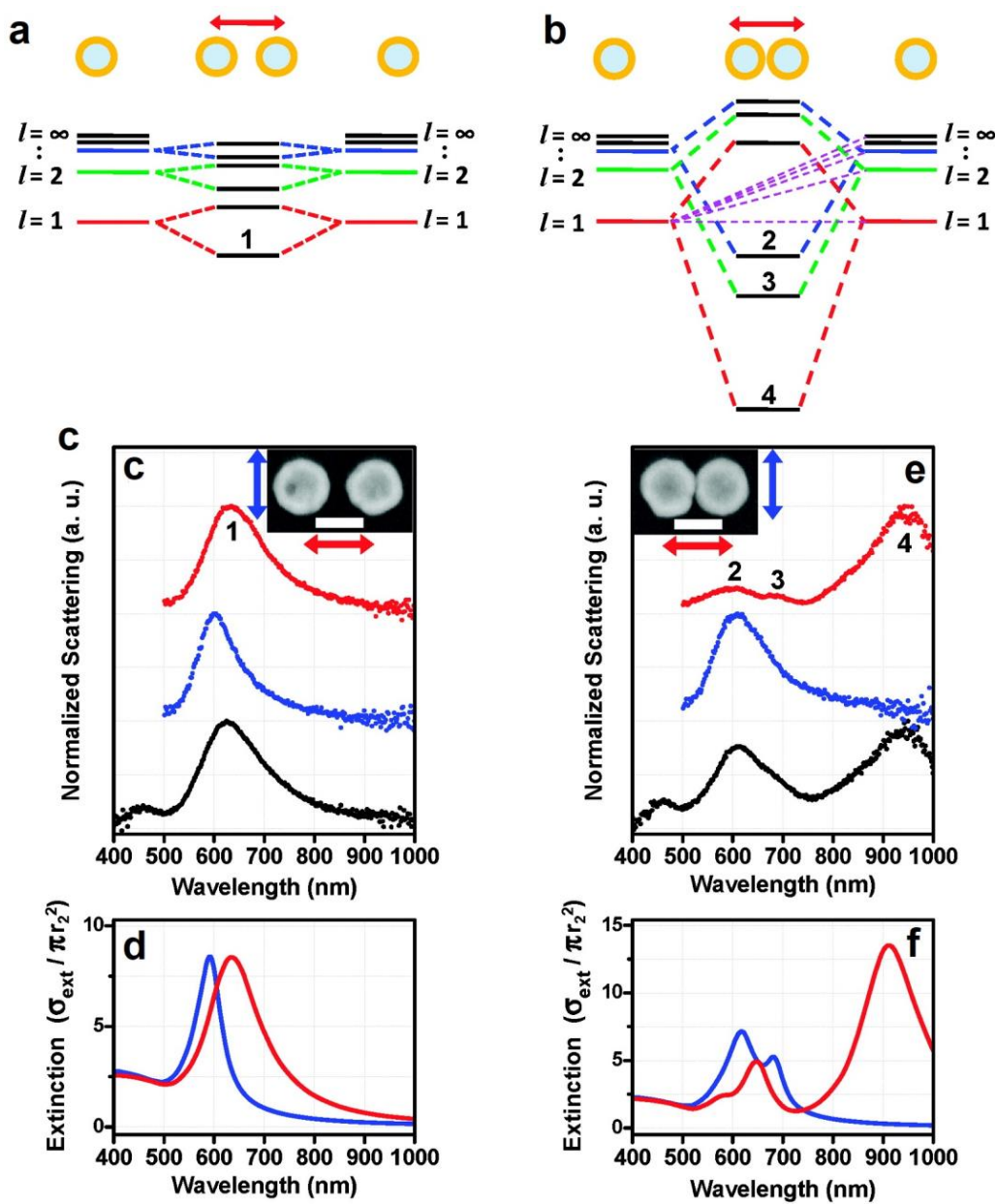

**Figure 4.** Energy level diagram for plasmon hybridization of nanoshell dimers, under *y*-polarized beam excitation for a) weakly, and b) strongly interacting dimers. Numbered energy levels correspond to numbered peaks in scattering profiles. c, d) Experimentally measured scattering response and simulation extinction spectra for a weakly interacting dimer, respectively. Both NPs were fit as $(r_1, r_2) = (42, 59)$ nm with the gap of 20 nm. Here $r_1$ is the inner and $r_2$ outer radii. e, f) Experimentally measured scattering response and simulation extinction spectra, respectively. Both shells were fit to be slightly elliptical, elongated in the transverse direction: the leftmost shell has a core with semiaxes 45 and 47 nm and an outer shell with semiaxes 58 and 60 nm, while the rightmost shell has a core with semiaxes 42 and 48 nm and an outer shell with semiaxes 55 and 61 nm with the gap of 1 nm. [178] Copyright 2008, American Chemical Society.

The experimentally and numerically studied extinction spectrum for the plasmonic dimer in weak coupling regime under both *x*- and *y*-polarized beams are plotted in Figures 4c and 4d, respectively. In this regime, the black spectrum represents illumination with unpolarized light while the blue (red) spectrum represents illumination that is polarized *y*-polarized (*x*-polarized) with respect to the dimer axis. The *x*-polarized wave radiation displays just a small red-shift of the dipole plasmon (peak 1) to the longer wavelengths. Noticing in both scattering and extinction panels, one can observe a small red-shift in the *x*-polarized dipolar plasmon extreme in respect to the *y*-polarized polarization spectral feature. In the strong coupling regime, the interaction between nanoshells is substantially stronger and the amount of red-shift is much more than the weak coupling system (Figures 4e and 4f). For the *x*-polarized illumination, the dipolar mode has red-shifted to the NIR band around ~930 nm due to the fact that the interaction mediated by the electromagnetic near-field, and also because the presence of the linker molecules between the nanoshells of the dimer structure. The consistency between experiments and simulations here was achieved by introducing a media with the RI of *n*=1.42, encompassing the nanodimer system. Furthermore, two higher energy modes are also excited in the *x*-polarized spectrum of the strongly coupled nanoshell, specified by peaks 2 and 3 (Figure 4e). These spectral features are correlating with the quadrupolar and octupolar plasmons in the dimer system, recognizing as the dark modes. Considering the spectral response of the structure in two different coupling regimes, once can reach to conclude that for large and classical distances between NPs, a strong bright superradiant mode can be observed. Conversely, in the strongly coupled dimer and hybridized condition, both dipolar bright and multipolar dark moments can be excited with high sensitivity to the dielectric variations of the media. However, the studied system above is quite simple holding broad resonances and cannot be employed for advanced and practical sensing technologies. To address this limitation, more complex nanostructures and multiparticle clusters such as quadrumers,[179] heptamers,[115,174] octamers,[180] and decamers[181] have been introduced, capable to support distinguished Fano-like and Fano resonances along a wide range of spectrum. Increasing the number of NPs in a cluster system and antisymmetric orientation of these particles lead to the formation of Fano resonance lineshape in the corresponding scattering response. The major reason behind the formation of Fano spectral feature is the destructive interference between the dark and bright modes in the energy continuum of the bright mode and emerging of a minimum.[180] Such a loss-less behavior and high sensitivity to the environmental perturbations have led researchers to employ these Fano-resonant subwavelength structures for biological and chemical sensing purposes.[179-182] On the other hand, Fano-resonant meta-atoms and metamolecules in periodic arrays,[123] known as metamaterials,[174,183-185] have been introduced as promising alternatives for the sensing purposes. In the following subsection we will consider the unique features and compare the performance of both technologies for practical sensing applications.

## 3.2. Fano-resonant clusters for biosensing

Categorizing as a unique spectral feature, Fano resonances can be distinctly appeared either in a simple single nanosphere[186] or a highly complex multipixel meta-atom.[183] Herein we provided a simple example for the excitation of Fano resonances in both multiparticle nanoclusters and multipixel unit cells. Liu and teammates have reported on the study of the excitation of multiple and pronounced Fano resonances in plasmonic heptamers consisting of split-rings along the NIR.[187] Figures 5a and 5b demonstrate the extinction cross-sections for a single split ring and sample seven-member heptamer cluster with the geometrical sizes, reported in the figure's caption. Setting the RI of the medium to $n$=1.33, two resonances are excited around ~955 and 1200 nm, correlating with the dipolar (bright) and quadrupolar (dark) moments, respectively (Figure 5a). Packing the split NPs in a symmetric fashion and launching *y*-polarized beam to the system lead to the excitation of Fano resonances due to the overlapping of quadrupolar mode with the bright dipolar mode (Figure 5b). Due to the presence of split sections in the rings, and because of the hybridization of optically driven plasmons, three Fano resonances are excited in the extinction profile. Running a quick 2D simulation allows to predict the influence of the RI variations of the medium on the spectral behavior of the nanostructure. Figure 5c illustrates the extinction spectra for the same structure with two different surrounding media conditions (air and water). Clearly, increasing the RI of the workplace substantially red-shifts the positions of all Fano dips to the longer wavelengths, which has been acknowledged as an inspiration for employing Fano-resonant structures for sensing applications.

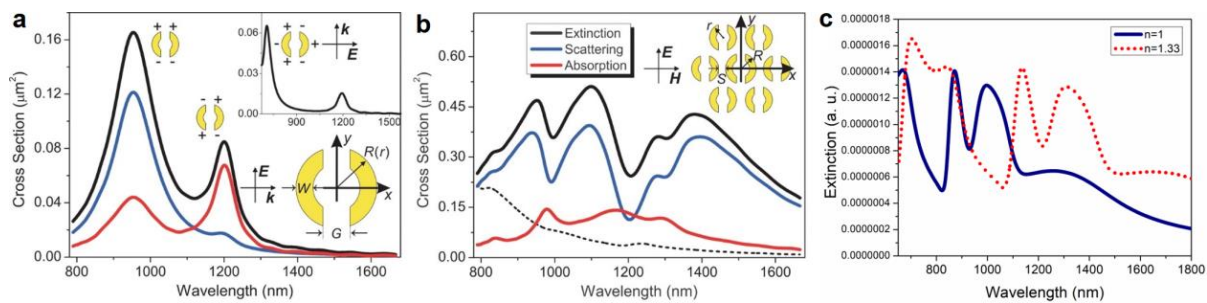

**Figure 5.** a) Spectra of a split NP under oblique incident illumination, where the parameters used are $W$=30 nm, $G$=30 nm, $R$=75 nm, the thickness $H$=40 nm, and the refractive index of surrounding medium is $n$=1.33. The inset on the upper right corner is the extinction spectrum with $x$-polarized beam. b) Spectra of a plasmonic heptamer cluster composed of SNRs, where the radius of center NP and surrounding particles are, respectively, $R$=80 nm and $r$=75 nm, the separation $S$=20 nm, and the dashed line is the extinction spectrum when the polarization is along the $x$-axis.[187] Copyright 2008, American Chemical Society. c) The extinction spectra for the same heptamer for two different RI of the medium.

In the following, we focus on the experimental applications of Fano-resonant nanoscale structures for the rapid and accurate sensing of ultra-small perturbations in the medium. To this end, we considered both RI sensing and biomolecule (bioprotein) sensing applications to provide a qualitative and comprehensive comparison at the end. Zeng and colleagues have proposed plasmonic Fano-resonant ultra-thin nanogratings integrated with a polydimethylsiloxane microfluidic channel for highly accurate RI sensing with extremely small variations in the media.[188] The authors claimed that the proposed integrated detection platform facilitates the RI resolution of around $1.46\times10^{-6}$ RIU accompanying with a high temporal resolution of 1 s. Figures 6a shows a prototype image of the proposed device integrated in a microfluidic channel, where the one-dimensional (1D) nanogratings with the thickness of 30 nm are deposited on a glass ($SiO_2$) substrate, as demonstrated in the scanning electron microscopy (SEM) image (Figure 6b). The measured and simulated spectral responses for the proposed structure are exhibited in Figures 6c and 6d, respectively. It is shown that for the gratings immersed in an aqueous medium, under transverse magnetic (TM) polarized beam illumination, two distinct asymmetric Fano spectral feature can be excited that are attributed to the constructive and destructive interferences between broadband LSPR modes in the nanoslits and two narrow distributing SPR on the top (Ag-water) or bottom (Ag-glass) metal/dielectric interfaces, respectively.[188] Considering the transmission amplitude spectra, slight increase in the transmitted intensity and the modulation percentage correlate with the strong SPRs excited by the beam transmitted across the nanoslits in the ultrathin metallic nanofilm. In this regime, the resonant amplitude of the SPRs is analogous to that of broadband DLSPRs, giving rising to a strong interference between these two modes and formation of pronounced multiple Fano minima. Figures 6e and 6f demonstrate the E-field intensity ($|E|^2$) plot at two different Fano lineshapes. In Figure 6e, one can see the excitation of LSPR moments around the edges of the gratings, at the glass-Ag interface, at the position of Fano dip 1, while at the other interface (water-Ag) the plasmonic resonant moments decay dramatically. On the other side, at the position of the Fano dip 2, the E-field distributions of LSPR and SPR moments are located at the water-Ag and glass-Ag interfaces, respectively. Thus the LSPR and SPR modes interfere with each other and produce robust asymmetric Fano resonances.

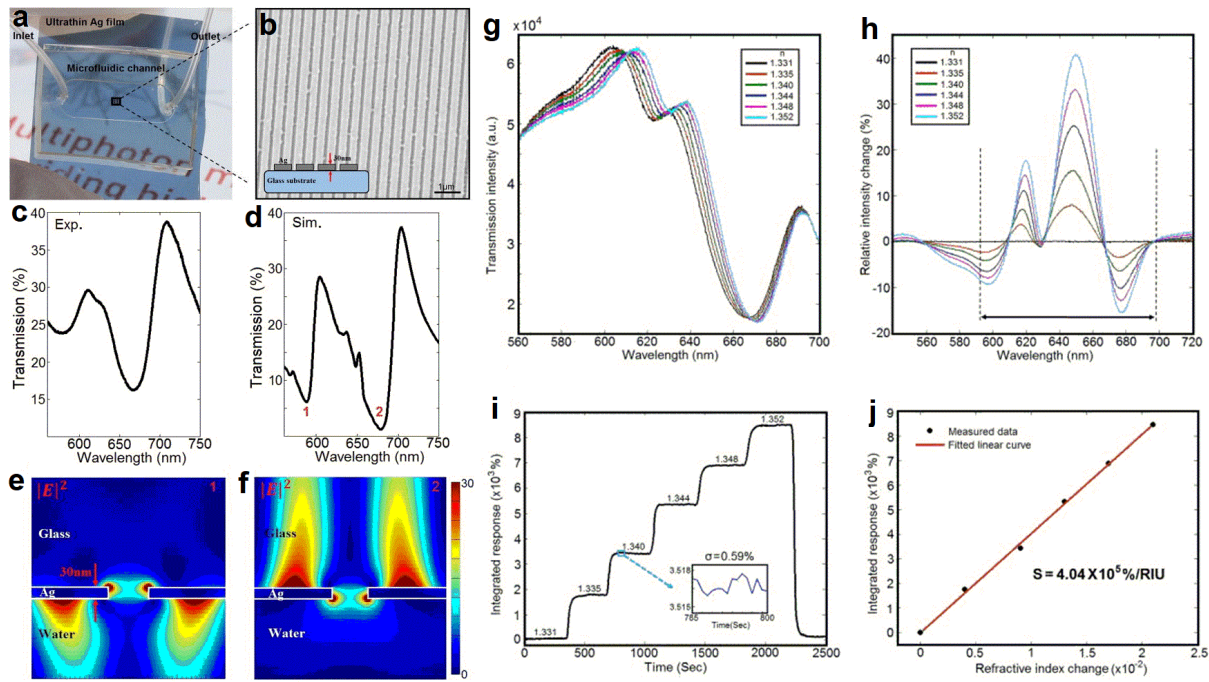

**Figure 6.** a) A photograph of an ultrathin 1D Ag nanogratings film integrated with a microfluidic channel. b) An SEM image of 1D nanogratings patterned on the ultrathin Ag film, with a period of 420 nm and slit-width of 110 nm. Inset is a cross-section of the ultrathin Ag nanogratings on a glass substrate. c, d) Experimentally and numerically obtained optical transmission spectra (under TM polarization excitation), respectively. e, f) Electric field intensity ($|E|^2$) distributions at two different Fano dip wavelengths, corresponding to transmission minima 1 and 2 in (**d**), respectively. g) The measured optical transmission spectra of the sensor in water and 3%, 6%, 9%, 12%, and 15% glycerol–water mixtures with RI ranging from 1.331 to 1.352. h) The relative intensity changes for liquids with different RI. The perpendicular dashed lines indicate the integration region. i) The integrated sensor response as a function of time. The inset shows the noise level of the sensor response over 15 data points with the time interval of 1 s. j) Extracted sensor output for liquids with different RI. The sloid line is the linear fit to the measured data (dots).[188] Copyright 2014, American Institute of Physics (AIP)

Figures 6g-6j exhibit the sensing performance and properties of the studied device. This was achieved by injection of glycerol–water solutions (0%, 3%, 6%, 9%, 12%, and 15% glycerol concentration) with measured RI ranging from 1.331 to 1.352 to the microfluidic channel. Figure 6g illustrates the transmission spectrum for the device for RI variations, reflecting a continuous red-shift in the position of the Fano dips to the longer wavelengths. The relative intensity changes percentage for liquids of different RIs are plotted in Figure 6h, showing signals with the highest prominence in the spectral range from 590 nm to 700 nm, as indicated by the black dashed lines. The magnitude of the relative intensity alterations was combined within this spectral range, providing the sensor output integrated response as a function of time (Figure 6i). The computed and measured integrated response values as a function of RI variations are plotted in Figure 6j. This shows that variation in integration response is approximately proportional to the increase in RI. The linear fit to the data points yield a sensor sensitivity of $4.04 \times 10^5$ %/RIU. The inset of Figure 6j indicates the noise level of the integrated response signal, which exhibits a standard deviation of 0.59% over 15 data points with a time interval of 1 s.

Next, we discuss and evaluate the sensing performance of advanced Fano-resonant nanostructures for the detection of biomolecules with low molecular weights at very small concentrations. In the plasmonic biosensors discipline, the role of the dark and Fano resonances modes cannot be neglected, and relatively promising, compact, label-free, and integrated nano- and microdevices have been developed and reported based on this technology.[128,174,182,185,189-192] In one of recent interesting studies, Liang and colleagues have tailored a plasmonic nanoplatform based on nanodisks located in nanoholes to excite subradiant dark resonances. This allowed them to design a sensing device for biomolecular detection based on the spectral response *via* strong binding of targeted proteins (Ribonuclease B (RNase B) and Concanavalin A (Con A) protein molecules) to the system.[193] Here the researchers have utilized RNase B bio-objects to catch and provide binding of Con A proteins. The molecular weight of RNase B and Con A are 13.7 kDa and 26.5 kDa, respectively, and detection of such small proteins using classical plasmonic systems is challenging. Hence, either dark modes or Fano resonances are proper choices for this purpose, however, the concentration of the proteins in the sensing media must be included in the evaluations. In the proposed approach by Liang *et al.*,[193] the examiners have exploited the subradiant dark mode arising from periodically fashioned hole-disk arrays under randomly polarized light (due to inherent symmetry), integrated and miniaturized with an optical fiber to probe biomolecules and proteins interactions, remotely and precisely (Figure 7a). This picture also contains the SEM image for the fabricated plasmonic surface. Similar to the previous study for Fano-sensing application, this device is also integrated with a flow cell. Figure 7b displays the normalized reflectance spectra for both simulation studies based on FDTD analysis and experimental measurements under polarized beam excitation. Visibly, two pronounced dips are induced at 750 nm ($\lambda 1$) and 838 nm ($\lambda 2$) with the quality-factor ($Q$-factor) of approximately 108 and 60, respectively. The E-field distribution over the top and bottom surfaces of the plasmonic arrays are illustrated in Figures 7c and 7d for $\lambda 1$ and $\lambda 2$, respectively. For the plasmonic mode located at $\lambda 1$, the E-field is mainly localized on the top surface of the structure, while the E-field is mainly localized at the bottom surface for the resonance positioned $\lambda 2$. This implies that two induced dips of the reflectance spectrum show both frequency and optical energy separations. Liang and teammates have utilized these advantages to develop a plasmonic biosensor.

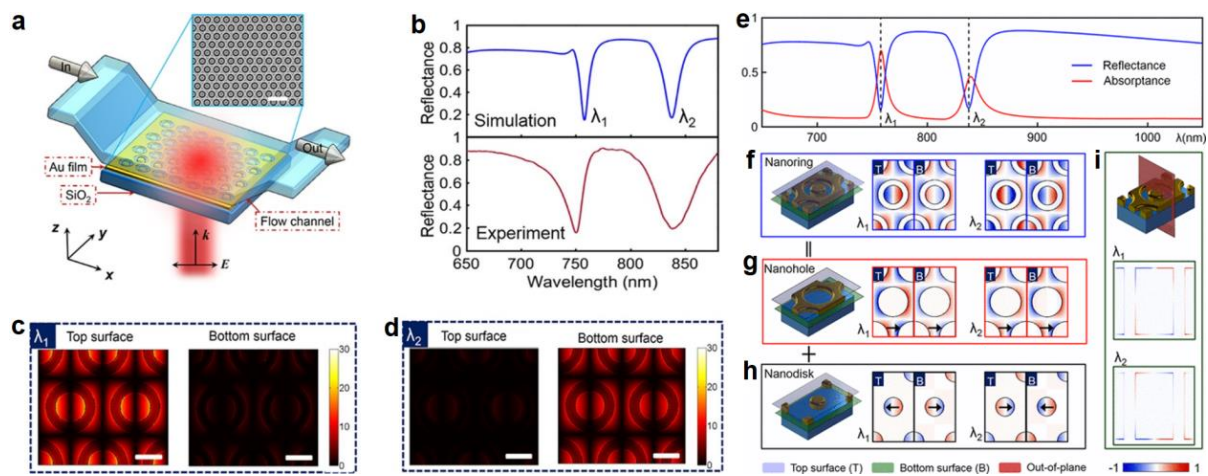

**Figure 7.** a) Schematic rendering of the miniaturized plasmonic sensor with a fluid flow channel and a top-down SEM image of the fabricated device (Scale bar, 2 μm). The periodicity of the hole-disk arrays is 640 nm and the thickness of the Au film is 100 nm. The inner radius (r) and outer radius (R) of the nanoring are 150 and 240 nm, respectively. b) The theoretically and experimentally obtained reflectance spectra. E-field distribution on the top and bottom surfaces of nanoring array at resonant wavelengths c) λ1 and d) λ2. e) Simulated reflectance (absorptance) spectrum of the plasmonic structure. The calculated in-plane normalized surface charge distributions of the relevant modes at both top and bottom surfaces of (f) nanoring, (g) nanohole, and (h) nanodisk arrays at resonant wavelengths λ1 and λ2, where the directions of dipolar modes are indicated by the arrows inside the maps. i) Numerically obtained out-of-plane normalized surface charge distributions of structure at resonant wavelengths λ1 and λ2. The top and bottom surfaces and out-of-plane are indicated by blue, green, and red planes, respectively.[193] Copyright 2017, American Chemical Society.

Describing by plasmon hybridization theory,[171] Figure 7e clearly shows the numerical results for both absorptance and reflectance spectra in a same diagram. The behaviour and nature of these modes can be qualitatively estimated and explained by plotting the surface change density maps for each component separately. Figures 7f-7i are the charge maps for the same system and simply considered each one of nanoring, nanohole, and nanodisk arrays at the position of two major wavelengths, respectively. Thus at $\lambda 1$, the dipolar modes at the top surfaces of nanohole and nanodisk show opposite distribution. In this limit, the dipolar momentum drastically damped, resulting in noticeable decline in the corresponding radiative losses similar to the dark modes,[138] which is originated from the antiparallel dipolar interference. The product of such a process is the efficient trapping of the incident beam and excitation of dark antibonding plasmonic resonances. The clue for this claim is the excitation of narrow lineshape at the position of $\lambda 1$. Given the fact that the dipolar modes correlating with the nanohole and nanodisks are oscillating in the same direction at $\lambda 1$, therefore, the dipolar interference strongly intensifies the total dipole, leading to a rapid depletion of the plasmon energy. Possessing stronger optical intensity at the top of the structure, the resonance at $\lambda 1$ much more sensitive to the surrounding dielectric variations, while the dipolar mode at $\lambda 2$ does not show the same sensitivity.[193]

In continue, we focus on the sensing properties of the investigated plasmonic device for the detection of targeted Con A bioproteins *via* binding with RNase B agents. As Liang and co-workers claimed, the tailored high performance biosensor can provide real-time detection *via* binding of RNase B and Con A protein molecules in solution on the sensing surface. Due to the having important role in most of organism studies associating with the cellular RNA degradation, the detection of low-molecular weight RNase B has received great interest in previous assays and examinations.[194,195] In addition, it is well-acknowledged that Con A protein holds benefits in applications needing solid-phase immobilization of glycoenzymes, especially the ones that have attested difficult to immobilize by conventional covalent coupling effect.[194-196] In the work presented by Liang *et al.*, the researchers have tried to capture the Con A biomolecule at different concentrations employing the proposed plasmonic sensor, functionalized with immobilized RNase B. This process is plotted schematically in Figure 8a. The specimens are prepared as follows: 1) the plasmonic chip is washed with pure water and ethanol, 2) then the substrate is immersed into an ethanol solution of 11-mercaptoundecanoic acid (MUA, 10 mM) at room temperature for 24 h to self-assemble an alkanethiol monolayer on the surface of Au layer, 3) afterward, the unreacted thiol biomolecules washed by utilizing ethanol, 4) once dried under N2 gas, a mixed liquidised solution containing 1-ethyl-3-(3-dimethylamino-propyl) carbodiimide hydrochloride (EDC, 0.55 M) and Nhydroxysuccinimide (NHS, 0.5 M) must be used for 30 min at 4 °C to activate the alkanethiol monolayer on the Au surface, 6) and finally, the biosensing chip is rinsed with deionized (DI) water and dried under N2 gas. By manipulating and integrating a microfluidic channel to the plasmonic chip, the researchers have injected 0.1 mg/mL RNase B in phosphate buffered saline (PBS) into the sensor microchannel for 30 min to form a stable monomolecular layer. Once rinsed with PBS buffer, bovine serum albumin (BSA, 0.1 mg/mL) is used for 30 min to deactivate the remaining activated carboxyl sites that did not combine with the RNase B. Carrying out set of experiments and characterization for different concentration of Con A bioproteins in the range of 0.02 mg/mL to 0.20 mg/mL, the shifts in the dark mode ($\lambda$1) position as a function of time and the association and dissociation rate between Con A and RNase B based on sensing curves have been monitored and plotted in Figure 8b. In this work, 5 s time resolution has been utilized to dynamically monitor the specific binding between RNase B and Con A proteins. As shown in Figure 8b, unexpected variations of the dark mode resonance after injecting Con A are observed in the detecting curves, which is ascribed to the possible occurrence of binding between Con A and RNase B molecules. It is well-accepted that reliability, LOD, and repeatability are the most important components to determine the performance of any optical nanosensor. These parameters have been determined by Liang *et al.*, where each concentration of Con A bioproteins is measured three times using the same senor chip (Figure 8c). Thus the LOD for the sensing of Con A is estimated to be 4.6 μg/mL (45 nM). Although the achieved results for the proposed nanostructure in this work are promising comparing to the classical NP-based assays, ELISA, RT-PCR, POCT, and electrochemical sensing methods, but the need for advanced biosensors with excellent LOD and accuracy stimulated scientists to find alternative approaches for optical biosensing purposes.

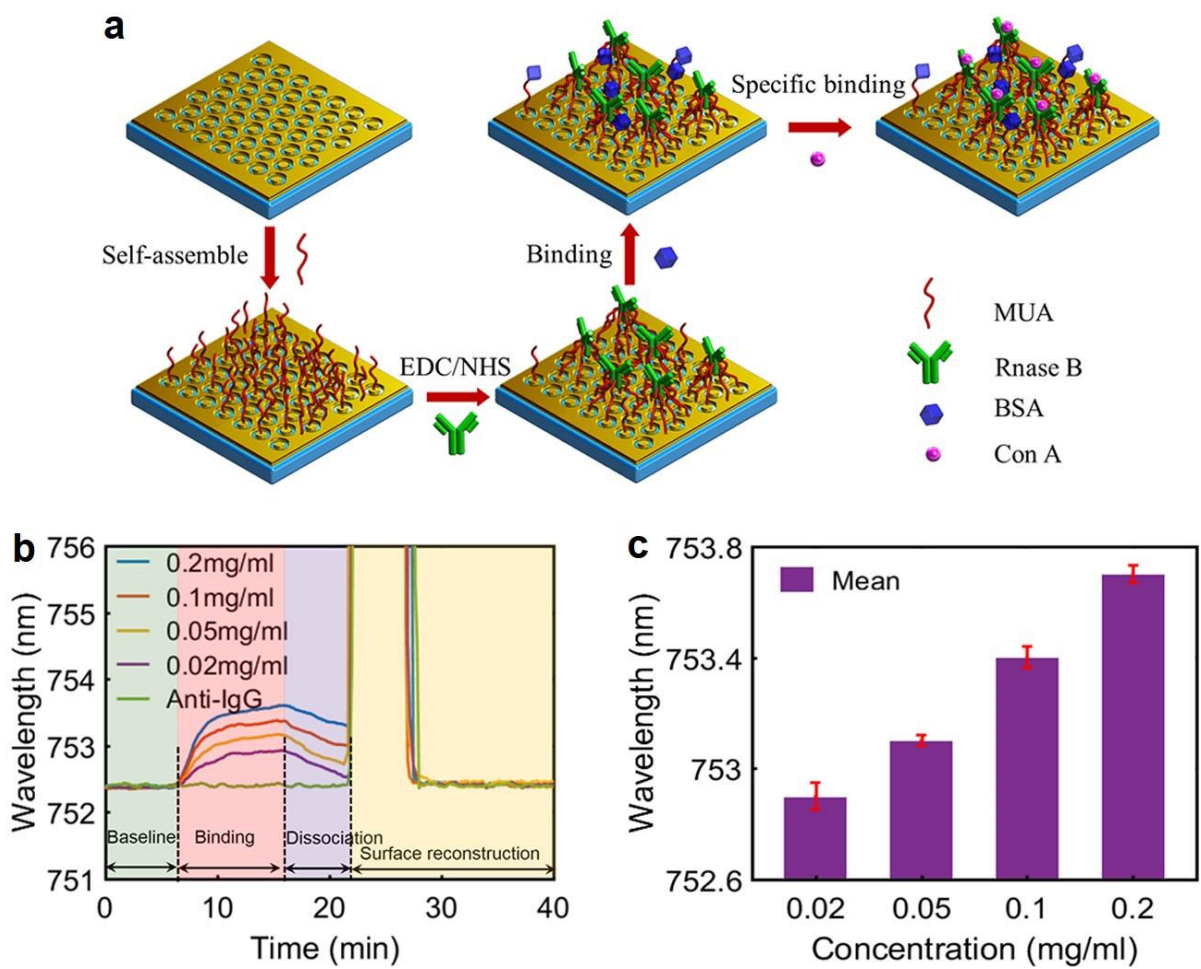

**Figure 8.** a) Experiment detection flow diagram for the binding of Con A and Rnase B biomolecules. b) Real-time biomolecular interaction response for the sensing of Con A at the concentrations of 0.02, 0.05, 0.1, and 0.2 mg/mL. The turquoise line illustrates the nonspecific binding of Rnase B and anti-IgG of 0.1 mg/mL. c) Relationship between the measured wavelength positions and Con A concentration.[193] Copyright 2017, American Chemical Society.

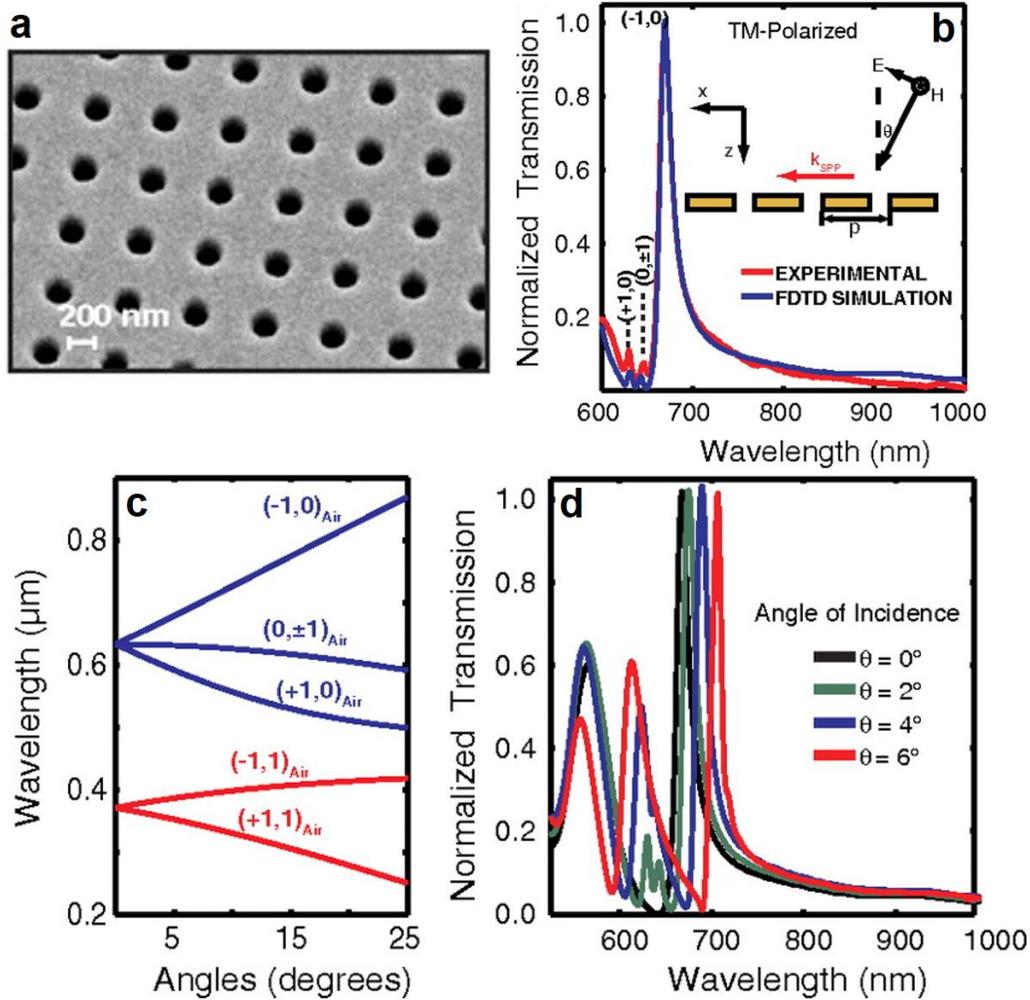

**Figure 9.** a) SEM image of the fabricated device based on interference lithography with the diameter and period of 230 nm and 580 nm, respectively. b) Experimentally (red curve) and numerically (blue curve) obtained transmission spectra. In the inset: the resonance peaks corresponding to the metal/air interface are indicated. c) Spectral dispersion of the transmission dip is presented as a function of the angle of the incident beam for several grating orders. d) Transmission spectra for varying angles of the incident beam using FDTD method analysis.[189] Copyright 2011, PNAS.

In addition to the previously reported Fano-based sensing methods, Yanik *et al*. and teammates have reported on the study of the spectacular and ultra-precise label-free observation and detection of IgG protein monolayers using high-$Q$ Fano-resonant nanoholes.[189] Figure 9a illustrates the SEM image for the fabricated nanoholes on an Au surface based on interferometric lithography technique, validating fabrication of a uniform device with negligible roughness. The experimentally measured and numerically (utilizing FDTD method) obtained transmission amplitude spectra are plotted in Figure 9b. This profile shows the formation of SPRs correlating with the different grating orders. Theoretically, for a given square lattice, the momentum matching condition between the in-plane wavevectors of the incident beam photons and the plasmon polaritons will be successfully met when the Bragg coupling condition is satisfied: $\vec{k}_{SPR} = \vec{k}_x \pm i\vec{G}_x \pm j\vec{G}_y$.[199,200] In this equation, $k_{SPR}$ is the surface-plasmon wavevecetor as: $|k_{SPR}| = (\omega/c)\sqrt{(\varepsilon_d \varepsilon_m)/(\varepsilon_d + \varepsilon_m)}$, where $c$ is the velocity of light in a vacuum, $\varepsilon_d$ and $\varepsilon_m$ are the permittivity of the dielectric and metal interface, respectively. In addition, $k_x$ incident light wavevector in $x$-direction, $\theta$ is the angle of the incident beam to the surface normal, $i$ and $j$ are the grating orders, and $G$ is reciprocal lattice vector order of the square lattice ($G=2\pi/p$, where $p$ is the periodicity of the gratings). Figure 9c illustrates the spectral dispersion of transmission dips as a function of the incident beam angle for different grating orders. Using this profile, one can perceive the dispersive trend of the resonant transmission peaks.[199-201] As clearly plotted in Figure 9c, at normal incident beam polarization, the transmission resonant modes associating with (0,±1), (+1,0), and (-1,0) grating orders at the metal-air interface are almost degenerated and showing polarization-independency. Yanik *et al.* have shown that due to the perpendicular trend of $k_x$ and $G_y$, hence, the transmission modes with the grating order of (0,±1) possess minimum dispersions for subtle variations in the angle of the illumination. For different illumination angles ($\theta \neq 0°$) and in the retardation limit, the antibonding modes (0,±1) and (+1,0) couple efficiently with the incident radiation (Figure 9d). The behaviour of dark modes at different angles and coupling of these modes to the radiations can be found in Ref. [189]. Here, we merely focus on the sensing capabilities of the proposed Fano-resonant device.

As it is discussed previously, the LOD of a given device can be defined by the corresponding linewidth of the induced resonant moments and their variations in the environmental RI perturbations. The RI sensitivity of the plasmonic nanoholes platform to the NaCl solutions with varying concentrations ($n$=1.3418, 1.3505, 1.3594, 1.3684) in DI water is demonstrated in Figures 10a-10c, obtained at room temperature. The insets are the transmission profiles as a function of energy for the induced dark and bright modes, fitted to the Fano interference model [189,201]:

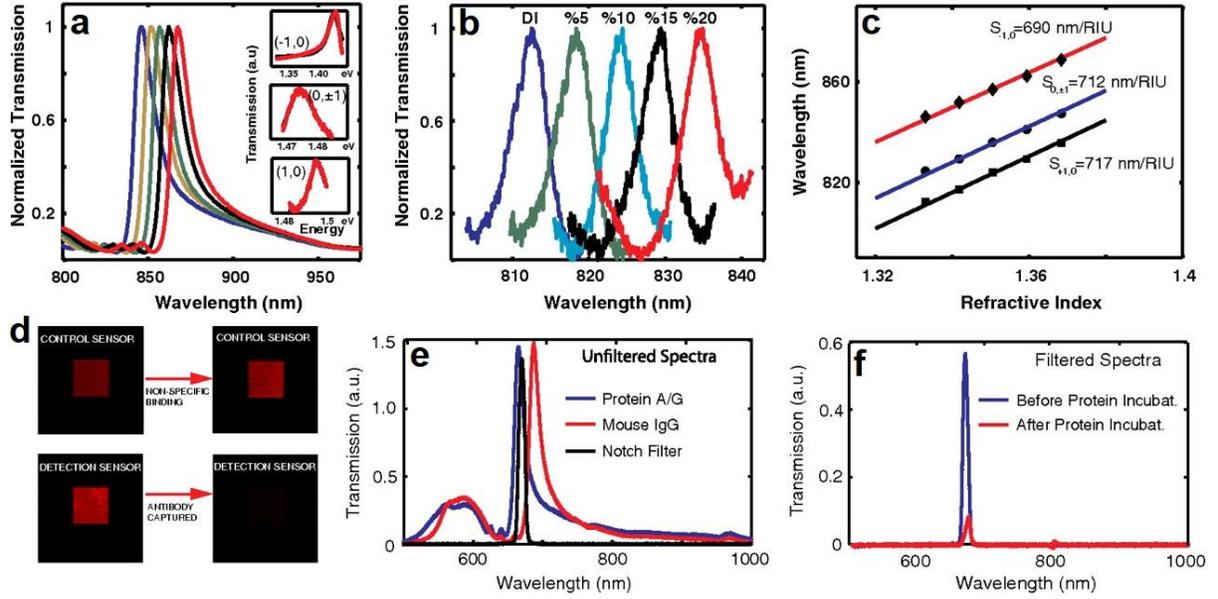

**Figure 10.** a) Spectral measurements profile for the variations in the concentrations of NaCl as a function of wavelength. b) Subradiant resonance lineshape (+1,0) shift is plotted for changes in NaCl concentration. c) Sensitivity of the devices are achieved using a linear fitting scheme. d) CCD images of the transmitted light obtained from detection and control sensors are evaluated. Capturing of the antibody causes a dramatic reduction of the transmitted light intensities through the detection sensors. e) Normali18zed transmission amplitude spectra are shown before (blue curve) and after (red curve) the capturing of the antibody. Spectral characteristic of the notch filter is also given (black curve). f) Transmitted light intensities in the presence of the notch filter is given before (blue curve) and after (red curve) the capturing of the antibody.[189] Copyright 2011, PNAS

$$T(\omega) = T_0 + A \left\{ \left(1 + \sum_v \frac{q_v}{\Delta \varepsilon_v}\right)^2 \Big/ \left[1 + \left(1 + \sum_v \frac{q_v}{\Delta \varepsilon_v}\right)^2\right] \right\} \quad (3)$$

where $\Delta \varepsilon_v$ is the transmission offset ($\Delta \varepsilon_v = 2\hbar(\omega - \omega_v)/\Gamma_v$), $A$ is the continuum-discrete coupling constant, $T_0$ is the transmission offset, $\omega_v$ is the resonant frequency, $\Gamma_v$ is the FWHM of the lineshape, and $q_v$ is the Berit-Wigner-Fano parameters, which defines the asymmetry of the lineshape for the *v*th resonance state. Accordingly, the highest RI sensitivity is obtained as 717 nm/RIU, obtained by utilizing high-*Q* Subradiant mode with the *Q*-factor of 196.9 (Figure 10c). In the end, sensing and seeing of biomolecular monolayers is explained by Yanik and colleagues. In an opposite method to the RI sensing that they employed the subradiant mode to see the bioproteins, the researchers here have used superradiant bright mode (-1,0) to see the proteins. It should be underlined that the bright mode has been used due to its coupling with the incident light, which is leading to the extraordinary light transmission. Additionally, A notch filter (FWHM≈ 10 nm) spectrally tuned to the plasmonic resonances peak is used to filter the light outside resonant transmission peak of the bright (-1,0) mode. The transmitted signal can be easily detected by naked eye without light isolation. Figure 10d illustrates CCD images of the transmitted light, in addition to images obtained from an unfunctionalized control sensor. Protein A/G (Pierce) is selected as the capturing biomolecule due to its high affinity to IgG immunoglobulins. Blue curve in Figure 10e shows the transmission spectra of a functionalized biosensor. By capturing of a single IgG antibody, 22 nm red-shift in the position of plasmonic moment is observed. This resonance shift is large enough to cause spectral overlapping of the transmission minima of the nanohole array with the transmission window of the notch filter (Figure 10f). Consequently, a dramatic reduction in transmitted light intensities is detected after the capturing of a monolayer of the antibody. This intensity variation is adequate to distinguish with naked eye in ordinary laboratory settings (Figure 10a). This detection technique not necessitating dark-ambience assays, exploits broad-band light sources allowing direct detection with human eye without any safety concerns. This mechanism of sensing offers exquisite advantages comparing to the previously analysed and recently reported Fano-enhanced methods, however, it is not addressing solutions for the detection of ultra-low concentrations of low-molecular weight and light biomolecules.

So far, there have been great efforts to show and validate how the dark-side of plasmons and Fano resonance mode can be used in much more accurate and efficient ways to detect biological objects and enzymes.[189,197,198] Table 1 lists the results for the latest achievements based on advanced plasmonic biosensing tools *via* Fano resonances and dark modes behavior. Obviously, in spite of providing substantial sensitivity, however, Fano-enhanced plasmonic biosensors show limited LOD, especially for ultra-low concentrations of low- molecular weight bio-agents. To address this shortcoming and to enhance the sensing quality of plasmonic devices, EIT-resonant meta-atoms and metamolecules have been accepted as promising tools for biological sensing applications. In the next section, we will consider and evaluate the recent achievements in this field of optical sensors.

**Table 1.** Comparison for the sensing performance of various biosensing plasmonic platforms.

| Description | Mechanism | Detected biomolecules & weight | S or FOM | LOD |
|---|---|---|---|---|
| Plasmonic nanoring-disk resonator arrays in visible [193] | Dark mode | Con A bioproteins, 26.5 kDa | 545 nm/RIU | 4.6 µg/mL |
| Ring-disk plasmonic cavities on conducting layer in visible [197] | Fano resonance | Immunoglobulin G (IgG), 50 kDa | 648 nm/RIU | ~1 mg/mL |
| Multipixel plasmonic metamaterials in MIR [198] | Fano resonance | Immunoglobulin G (IgG), 50 kDa | N/A | ~1 mg/mL |
| Plasmonic nanohole arrays in visible [189] | Fano resonance | Immunoglobulin G (IgG), 50 kDa | 717 nm/RIU | ~1 mg/mL |

## 4. Perfect plasmonic metamaterial absorbers for sensing

Perfect metamaterial absorbers are artificially tailored subwavelength structures consisting of periodic arrays of unit cells, offering unique advantages by utilizing the inherent lossy behavior of plasmons.[202,203] Perfect absorption of incident beam has been achieved *via* developing well-engineered plasmonic meta-atoms and metamolecules across a wide range of spectrum.[203,204] In terms of optical physics, metamaterials can be described using a complex electric permittivity and magnetic permeability as follows:[205]

$$\begin{cases} \tilde{\varepsilon}(\omega) = \varepsilon_1 + i\varepsilon_2 \\ \tilde{\mu}(\omega) = \mu_1 + i\mu_2 \end{cases} \quad (4)$$

Most of the research in the development of advanced metasurfaces has been centered on the real parts of permittivity and permeability components, enables the formation of optical structures with negative-RI material.[206,207] Although this strategy have been employed for developing advanced plasmonic and all dielectric devices from radio frequencies (RF) to the ultraviolet (UV) spectrum, however, the lossy (imaginary) part of the components in Eqn. 4 have potential power for the creation of new structures with advanced properties. Perfect plasmonic absorbers are one of promising applications of this technique, which can be tailored *via* the independent manipulation of resonances in ε and *μ*. This leads to the absorption of both components of the incident electromagnetic field by minimizing the reflectance spectral feature. This technology has been extensively employed recently to address conventional limitations correlating with the biochemical and biological sensing devices. In the following subsections, we first introduce the physical mechanism behind the formation of perfect absorption of incident beam and later the recent progresses in the RI and bio sensing assays will be explained.

### 4.1. Perfect plasmonic metamaterial absorbers

In this section, we analyze the theory of the perfect absorption feature in metamaterials by providing an example, published by Landy and colleagues.[202] Figures 11a-11c show schematic and geometrical description for the studied plasmonic multipixel unit cell composed of two resonators to support electric and magnetic resonances. This would allow to absorb all of the incident electromagnetic field at the corresponding resonant frequency. The resonator in Figure 11a supplies the electric coupling and the magnetic coupling is obtained by using the cut wire resonator shown in Figure 11b. These resonators are splitted with a capacitive waveguide as a spacer, which possesses critical role in tuning the magnetic resonance. Accordingly, for the metamolecule with the following geometries: $a_1$=4.2 mm, $a_2$=12 mm, $W$=3.9, $G$=0.606 mm, $t$=0.6 mm, $L$=1.7 mm, $H$=11.8 mm, where the resonators are separated with 0.65 mm capacitive waveguide. Figure 11d illustrates the numerically obtained transmission and reflection amplitudes as a function of incident beam frequency. Clearly, almost unity absorption is achieved around 11.56 GHz, which obtained by tuning the frequency of the permittivity and permeability components to satisfy $\varepsilon=\mu$ (leading to an impedance near the free space value). Here the obtained absorption ($A$) is approximately 100%, the reflection ($R$) and transmission ($T$) are 5% and ~97%, respectively. Thus one can use the following classical equation to define the absorption: $A$=1-$T$-$R$, around 99%, with a FWHM of 4% compared to the induced resonance linewidth. Taking the advantage of the sharp lineshape of perfect absorption feature, one can provide highly sensitive plasmonic sensors. Furthermore, the experimental measurements for the reflection and transmission spectral responses are plotted and compared with simulation results in Figures 11e and 11f, respectively. The exceptional fashion of the metamolecule leads to having strong sensitivity to the polarization direction of the incidence, hence, this behavior can be used for various applications such as switching and modulation.[208,209] However, in this work we mainly focus on the potential sensing applications and properties of analogous integrated subwavelength structures in much shorter wavelengths with smaller dimensions.

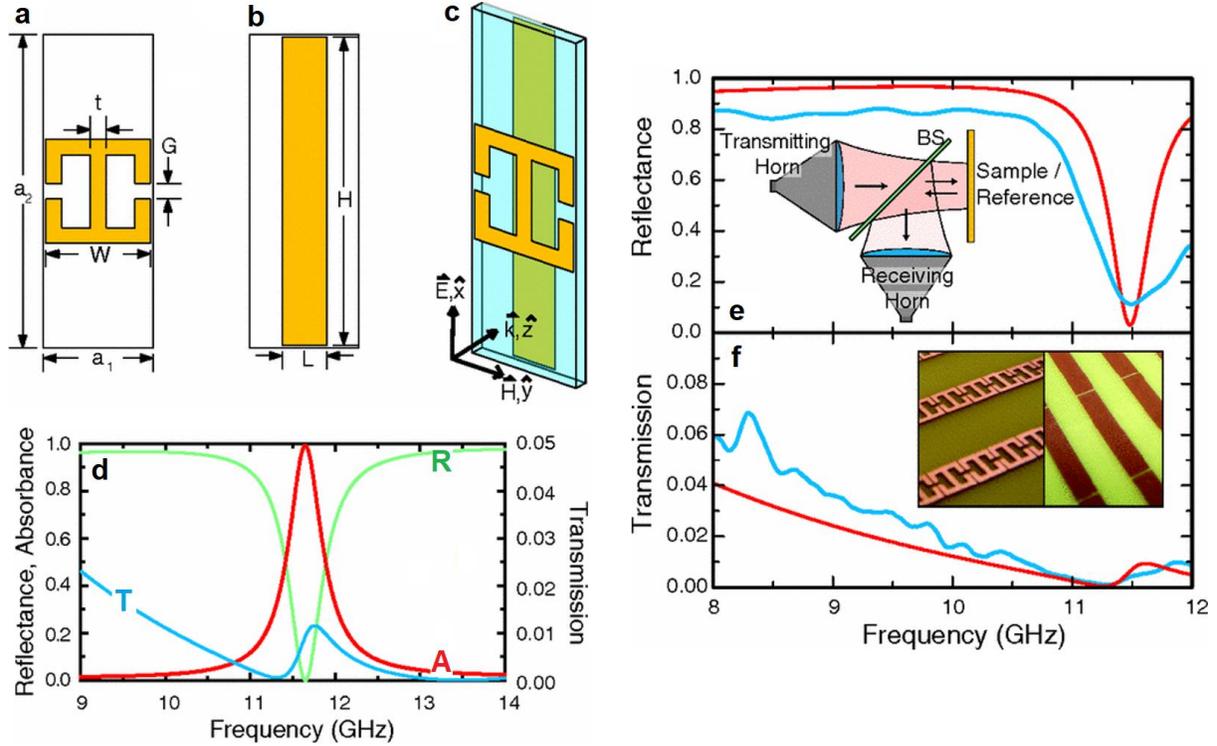

**Figure 11.** a) Electric and b) cut wire magnetic resonators. Dimension notations are listed in (a) and (b). c) The unit cell with axes indicating the propagation direction. d) Reflection, absorption, and transmission amplitude spectra. e) and f) the simulated and measured reflection and transmission spectra are plotted, respectively. The insets are the microscopic images of the fabricated metamaterial.[202] Copyright 2008, American Physical Society.

### 4.2. Perfect plasmonic metamaterial absorbers for refractive index sensing

In previous sections, we comprehensively discussed the importance of the RI sensing, which is the initial and fundamental technique to estimate the sensitivity of biological sensor and transducers.[181,210] Similar to previously reported Fano-based and classical plasmonic biochemical sensors, perfect plasmonic metamaterial absorbers also follow the same mechanism of sensing, however, they show highly remarkable sensitivity to the medium dielectric value variations. As one of leading works in this concept, Liu *et al.*[203] have numerically and experimentally validated and shown that perfect plasmonic metasurface absorbers are extremely sensitive to the minor RI variations in the surrounding media. By introducing periodic arrays of subwavelength metallic disks on a multilayer functional surface consisting of $MgF_2$, Au, and glass layers, the researchers successfully showed that the metasurface is able to support perfect absorption around NIR of spectrum (Figure 12a). In the polarization-independent absorber sensor, the Au sublayer acts as a mirror to eliminate the transmission of the beam ($T=0$), and thin $MgF_2$ layer (with the permittivity of 1.9) is the dielectric spaces between metallic nanodisk and sublayer. The reflectance profile for the proposed structure is simulated and plotted in Figure 12b, where the spectra is computed for three different damping constant of bulk Au. According to Drude model,[211] the plasma frequency ($\omega_p$) and damping constant ($\omega_c$) of Au can be defined and tuned to achieve the best results. Here by setting the following values: $\omega_p=1.3\times10^{16}$ s$^{-1}$, and $\omega_c=4.08\times10^{13}$ s$^{-1}$ one can see the best reflectance obtained for three time damping constant of bulk Au.

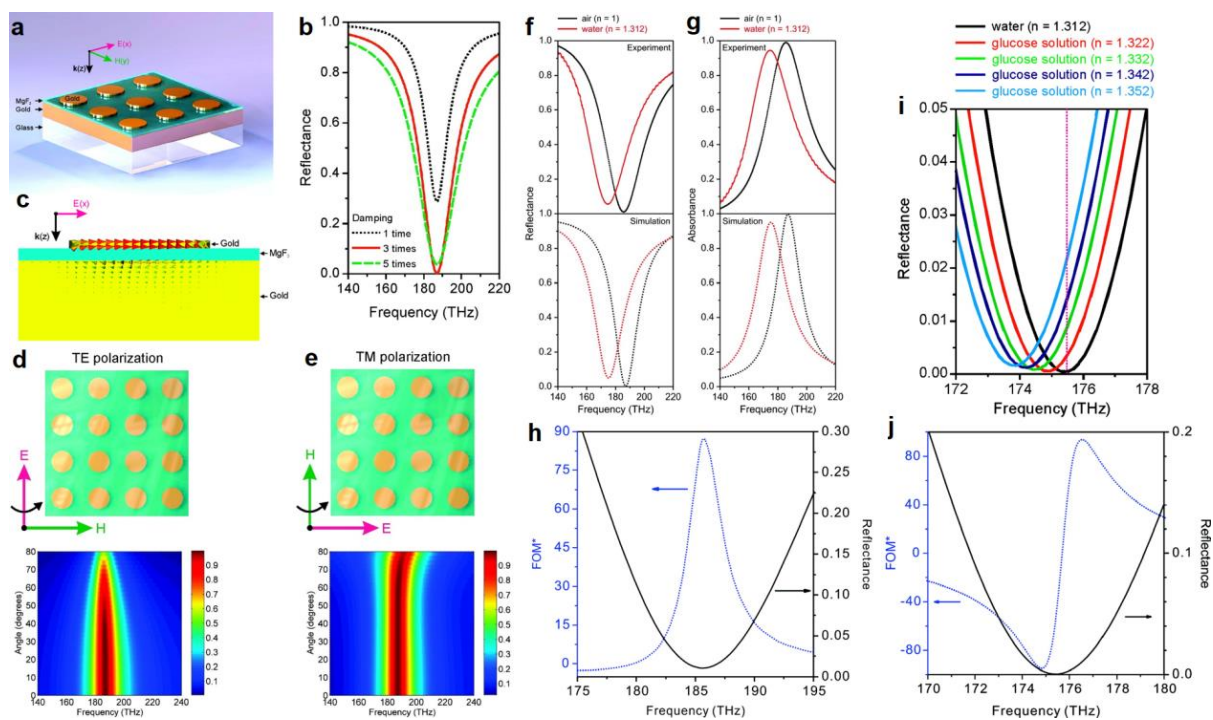

**Figure 12.** a) 3D rendering of the perfect plasmonic metamaterial absorber structure. The diameter and thickness of the Au disks are 352 and 20 nm, respectively, located with the periodicity of 600nm in both *x* and *y*-directions. The thickness of the MgF$_2$ spacer is 30 nm and the thickness of the Au mirror is 200 nm. The whole structure resides on a glass substrate. b) Numerically computed reflectance spectra in for the variations in the damping constant of the Au film. c) Calculated current distribution at resonance where perfect absorbance occurs. Antiparallel currents are excited in the gold disk and the gold film. d, e) The simulated angular dispersions of the absorbance peak for TE and TM configurations, respectively. f, g) The simulated reflectance and absorbance spectra with different dielectric materials (air and water) on the structure surface, respectively. h) Experimental FOM as a function of frequency. The highest value of FOM is obtained as 87, which is reached at a frequency slightly away from the minimum reflectance. The experimental reflectance spectrum with air on the structure surface is replotted with a black curve. i) Simulated reflectance spectra of an absorber sensor designed for water as reference medium. Spectral tuning occurs with glucose solutions varying from 0 to 25% that have different refractive indices. j) Calculated FOM as a function of frequency. The maximum value is 94, and it is reached slightly away from the reflectance minimum. The simulated reflectance spectrum with water on the structure surface is replotted with a black curve.[203] Copyright 2011, American Chemical Society.

Figure 12c allows to understand the operating mechanism of the absorber tool. The current distribution profile shows the formation of antiparallel currents that are excited in the surface nanodisks along the metallic sublayer, known as magnetic resonance, interacting strongly with the magnetic component of the incident beam.[212] In this regime and at the resonance frequency, due to the strong field enhancement between the metallic disks and sublayer, a huge energy can be confined at the capacitive spacer, which blocks the reflection of light. Thus one can claim that the absorption is almost unity and around 100%. Figures 12d and 12e demonstrate the angular dispersions if the absorption extreme at several angles of the illumination for TE and TM polarizations, respectively. As obvious in the corresponding snapshots, for the TM polarization, the absorption peak is independent of the angle of incidence due to the fact that the direction of the magnetic field of the incidence remains fixed for varying incident angles, while for the TE polarized beam, the magnetic field cannot trig the circulating currents efficiently at large angles, hence the absorption reduces drastically.

Next, we focus on the sensing performance of the analyzed multilayer perfect absorber metadevice. Firstly, the spectral behavior of the reflection and absorption curves to two distinct environments (air and water) is studied, as exhibited in Figure 12f and 12g, respectively, for both experimental and simulation analyses. Conventionally, by increasing the RI from $n=1$ to $n=1.312$, major red-shifts in the position of both reflection and absorption resonances have been monitored from 175.5 THz to 185.6 THz. Therefore, the corresponding FOM can be computed,[203,213] using the following equation:

$$FOM = Max \left| \frac{dI(\lambda)/dn(\lambda)}{I(\lambda)} \right| \qquad (5)$$

where $dI(\lambda)/dn(\lambda)$ is the relative intensity change at a given resonance wavelength excited by variations in the RI ($dn$), and $I(\lambda)$ is the intensity of the peak, where the FOM reaches its maximum value. The experimental FOM characteristic profile is plotted in Figure 12h, and for $dn=0.312$, the peak os FOM is located as 87 at the perfect absorption frequency. The solid curve in this diagram represents the reflection spectrum. The most important advantage of the studied absorber sensor by Liu and teammate lies in the fact that it allows recognition of photons that are reflected by the nonperfect absorber upon RI change against a nearly dark reference measurement, where only few photons are reflected from the perfect metasurface absorber. The corresponding sensitivity of the absorber sensors to the subtle variations is quantified as 400 nm/RIU. To show the strong sensitivity of the perfect absorber structure to the ultra-small variations in the RI of the medium, glucose solution with varying RI (ranging between 0 to 25%) is employed and the simulation results for the variations in the reflection spectrum are demonstrated in Figure 4i. A continuous red-shift in the position of the magnetic resonant moment is observed for variations in the glucose solution RI, which helps to accurately quantify the sensitivity of the plasmonic sensor. The maximum value for FOM is obtained as 94. The simulated reflectance spectrum with water on the structure surface is replotted with a black curve. As it is discussed, the flexible plasmonic metamaterial absorber provides significant absorption cross-section across the NIR and also shows super-sensitivity to the surrounding environment RI variations. This mechanism is a practical use of the lossy behavior of plasmonic structures and using the undesired aspect of plasmons in an efficient and practical way. In the upcoming subsection, we describe the possibility of utilizing perfect metasurface absorbers for the detection of biomolecules at low concentrations.

## 4.2. Perfect plasmonic metamaterial absorbers for biological detection

In this part, we study and briefly review the recent achievements in the field of perfectly absorptive plasmonic metamaterials for label-free and practical bioprotein and biomolecular detection. Among all of the recent work for the declared biosensing purpose, as a growing trend, metallic NPs-integrated perfect metasurface absorbers have received copious attention.[214,215] It is shown that the integration between nanosize metallic NPs and microsize plasmonic structures leads to developing precise, cost-effective, integrated, and label-free plasmonic biological sensors. Very recently, Xu *et al.*[215] have shown that perfect metasurface absorbers can be tailored to provide quite high absorption profile along the terahertz (THz) band. Integration of Au spherical NPs with microscale plasmonic unit cells helps to use both the high absorption and strong binding of biomolecules corresponding with the absorptive metamolecules and NPs, respectively, simultaneously. The use of THz metasurfaces for biological and pharmacology applications originates from their promising features including but not limited to quick infection diagnosis, cost-effective and real-time analyses, owing to their non-destructive and harmless interaction with biological tissues in both *in vivo* and *in vitro* assays.[216,217] The THz spectra, indeed, is a gap between the laser and electronic worlds, majorly lies in the following frequency range: $0.1 \text{ THz} \leq f \leq 10 \text{ THz}$.[218-220] The enhancement of electromagnetic field near the metallic metamolecules empowers the tailored devices to sense trace concentrations of targeted biological agents, antibodies, bacteria, proteins, etc.[215,221,222] Despite of offering unique and promising advantages, THz metasurfaces reflect poor performance for detection of microorganisms that are non-responsive and transparent to the incident beam in this spectrum. This shortcoming originates from the low scattering cross-section for the micro and nanoorganisms with size of $\lambda/100$.[221] Numerous strategies have been carried out to overcome this drawback such as inducing ultra-narrow resonance lineshapes, and using the plasmonic behavior of 2D materials (i.e. graphene), etc.[223-225] In addition to these solutions, integration of plasmonic nanosize particles and THz metamolecules has been acknowledged as a most promising technique to enhance the sensitivity of developed devices.[215] The key benefit of this method robust binding between metallic NPs and bio-targets, leading to the detection of targeted molecules at very low concentrations.

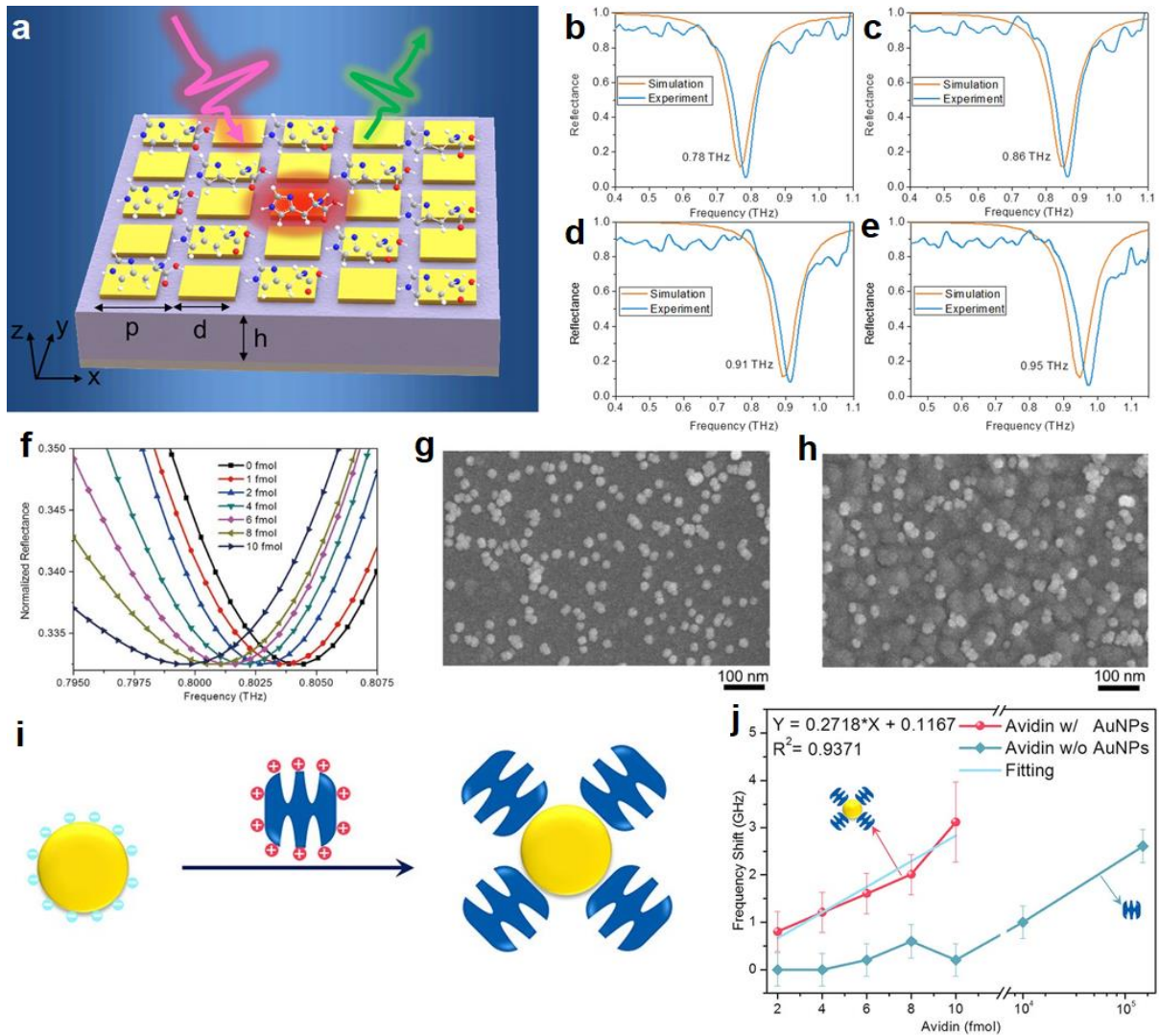

**Figure 13.** a) 3D diagram of plasmonic THz metasurface. Simulated and experimental reflectance profiles for the length of the metal patch as follows: b) 105 μm, c) 95 μm, d) 90 μm, and e) 85 μm. f) Reflective curves resulting from different amounts of gold NPs. g, h) SEM images of 100 fmol Au nanoparticles on photoresist and an Au patch, respectively. i) Principle of conjugating avidin and gold nanoparticles. j) Frequency shifts resulting from avidin and conjugated avidin−Au NPs.[215] Copyright 2011, American Chemical Society.

As one of recent and pioneer works in the field of biosensing, Xu and colleagues have shown that blending highly absorptive metamolecules and NPs with each other leads to the recognition of Avidin protein (a tetrameric biotin-binding protein) with an estimated molecular weight of 66-69 KDa.[226] Here, we briefly review the mechanism has been utilized to detect the presence of such small proteins at the concentration of approximately 100 pmol. The perspective representation of the proposed metamaterial is demonstrated in Figure 13a, consisting of arrays of metallic blocks to support spoof-type plasmons with sharp and deep reflection profile similar to perfect absorbers. The engineered metasurface reduces the intensity of the plasmons at the resonances through damping effect (changing beam to heat),[4] giving rising to formation of enhanced absorption peak. The reflectance spectra for both simulation and experimental studies are plotted in Figures 13b-13e for different length of the metallic blocks, ranging between 85 µm to 105 µm. The strongly localized enhancement of the electromagnetic field coupled with the metasurface enables a new signal-enhancement technology by measuring the resonance frequency shift caused by the sensing targets. Here the resonance frequencies of the metamaterials increase with decreasing the metal patch size. The other important feature of this metasurface is the magnetic and the electric resonance field enhancements of the impedance matched absorption cavity, which enable stronger interactions with the dielectric analyte for the same pattern of metamaterial absorber, enhancing the sensitivity of the structure substantially.[227] To show how the plasmonic metastructure is sensitive to the environmental variations, Xu *et al*. have examined the spectral response of the sensor and red-shift of the absorption peak, which is presented in Ref. [215] in details.

In continue, by introducing various concentrations of Au NPs (ranging from 1 to 10 fmol) to the metamaterial system (transparent to the THz beam), a subtle red-shift in the position of the reflectance dip is observed and demonstrated in Figure 13f. This shift is due the concentration of NPs that affects the electromagnetic field slightly. For instance, for 10 fmol concentration of Au NPs, the frequency shift reached 4.22 GHz. Figures 13h and 13g are the SEM images of 100 fmol Au NPs on photoresist and an Au patch, respectively. It should be underlined that the photoresist (SU-8) layer is used as a dielectric layer in the middle, and Au is used as a metallic layer on the top in the system. The presence of Au NPs helps to increase the dielectric permittivity of the surrounding media, which improving the sensitivity of the system by a few fmols. As a proof of concept, avidin and conjugated avidin−Au NPs ranging from 2 to 10 fmol are introduced to the metamaterial. Figure 13i illustrates the principle of conjugating avidin and Au NPs schematically. In this study, each Au NP (with an approximate diameter of ∼10 nm) is negatively charged and coated with citrate, while avidin (with a diameter of ∼7 nm) is positively charged with a given pH (∼7) that is lower than its isoelectric point (pH = 10). Thus, electrostatics can account for the strong binding between and conjugation of avidin and Au NPs. Considering the interspace interference and the electrostatic repulsion between proximal avidin proteins, thereby the proportion of avidin molecules to Au NPs will be no larger than 15:1. The corresponding resonance frequency shifts as a function of avidin protein concentration (conjugated with Au NPs) is exhibited in Figure 13j. Accordingly, for the absence of Au NPs and direct binding between avidin and microsize patches, one can observe trivial shift in the position of absorption peak. Conversely, in the conjugated avidin− Au NPs limit, the frequency shifts showed reasonable linearity with the amount of Au NPs concentration, yielding determination coefficient of $R^2 =$ 0.9371. The LOD for the plasmonic metamaterials with the conjugated avidin−Au NPs is determined as 7.8 fmol. For instance, for 150 pmol avidin, a frequency shift of 2.61 GHz observed, where for 10 fmol conjugated avidin−Au NPs, the frequency shift attained up to 3.12 GHz, showing more than a 1000-fold enhancement in the corresponding sensitivity. Hence, the introduction of Au NPs has the potential to provide a strong enhancement in the detection sensitivity of THz metamaterials.

By far, we comprehensively reviewed the recent and previously reported achievements in the use of plasmonic nano and microplatforms for developing advanced biosensing tools with high precision. Playing a promising role in the modern and next-generation clinical and medical techniques, plasmonic metasensors technology is still under development. The major shortcoming correlating with plasmonic sensors components that must be addressed are sensitivity, selectivity, and repeatability. Despite the enormous amount of progress in this field of practical optical devices, the mentioned factors have not been enhanced significantly. This originates from the inherent limitations in the induced classical electromagnetic moments in plasmonic platforms, introduced by Maxwell and Lorentz and later cultured by Landau and Jackson.[5,186,228] Therefore, to overcome these limitations, the classical behavior of plasmonic structures must be modified or enhanced. In the next section, we show the recently introduced and analyzed alternative platforms for conventional plasmonic sensors.

## 5. Toroidal resonances for sensing

Toroidal multipoles have been categorized as a third family of multipoles, apart from classical electromagnetic multipoles, introduced by Zel'dovich,[137,229] and studies for the first time in the context of classical electrodynamics,[230] solid state physics,[230,231] and atomic, nuclear, molecular physics.[232,233] In recent years by developing 3D and planar metamaterials (operating along microwave frequencies along the visible band), a comprehensive theoretical model has been established for toroidal electrodynamics.[234,235] Toroidal dipole is the fundamental member of toroidal contributions which is a configuration of static currents flowing across the surface of torus, has initially been observed in nuclear physics, known as "static toroidal dipole" or "anapole moment".[231,233,236,237] Conversely, the dynamic toroidal dipole (optically driven mode) is an independent term in the family of electrodynamic multipole expansion.[237,238] In terms of the condensed matter principle, theoretically, the dynamic toroidal dipole is the direct residues from an electric octupole and a magnetic quadrupole, can be combined to form a squeezed nonradiating head-to-tail charge-current configuration, known as an axial toroidal dipole moment (**T**).[234,239] In terms of the optical physics, time reversal ($t \to -t$) and space inversion ($r \to -r$) symmetries are the fundamentals of the toroidization principle for both vortex-like and polar charge-current configurations.[237] Such violations in the classical electromagnetic symmetries give rise to the formation of intensely squeezed vortex of a closed-loop arrangement with ultra-weak far-field radiation signature, which makes toroidal moment as a "*ghost* *resonance*" (Figure 14a).[240,241] The main result of this interference is a nonzero contribution to the scattered far-field radiation, which can be characterized principally by the summation of induced charge, and induced transversal and radial currents. The toroidal dipole is a strategic member of toroidal electrodynamics due to having undeniable electric field radiation pattern, which possesses much more sensitivity to the environmental perturbations in comparison to the classical resonant moments. The following equations compare the electric field radiated from a traditional dipole and toroidal dipole:

$$\begin{cases} E_p = n^2 k_0^2 \left( \hat{\mathbf{r}} \times \hat{\mathbf{r}} \times \dfrac{p}{r} \right) \\ E_T = n^3 k_0^3 \left( \hat{\mathbf{r}} \times \hat{\mathbf{r}} \times \dfrac{T}{r} \right) \end{cases} \qquad (6)$$

in which **r** is the vector linking the location of the dipolar moment with the observer, and *n* is the RI of the media. As key equations, clearly, in the toroidal moment limit, controlled variations in the dielectric characteristics of the surrounding ambience leads to much more variations in the scattered electric field radiation. Such a spectral feature facilitates tailoring highly precise plasmonic sensors with excellent sensitivity to the RI alterations and also the presence of ultra-low-weight biomolecules, proteins and antibodies. In the following subsection, we briefly describe the theory behind the excitation of dynamic toroidal dipoles in plasmonic metamolecules and meta-atoms. Additionally, the recent progress in the field of toroidal metasensors and their exquisite advantages will be explained.

## 5.1. Theory of toroidal multipoles as ghost resonances

Previously, it was assumed that the toroidal transition matrix elements tend to be weak for atoms and small molecules, hence, the contribution of toroidal dipole to the electromagnetic response of matter was ignored.[136,242] It is well-accepted that metamaterials are promising platforms to support both classical electromagnetic and toroidal multipoles and near-field characteristics.[237,238,243-245] Because of the inherent complexity of the charge-current distributions that excites the toroidal dipole moment, such a mode can merely induce in well-engineered subwavelength scatterers.[135,136] The initial theory for the formation of toroidal multipoles have been described by monitoring the radiation from an infinite planar arrays of given subwavelength metamolecules.[136,246] It should be underlined that concealing the far-field radiation pattern hinders easy detection of toroidal radiation signature, hence, the far-field distribution of the electric field radiated from a single oscillating toroidal dipole must be considered as a first step to model the theory. In 2002, Radescu *et al.*[247] have shown that the electric field emitted from a random scatterer can be derived by:

$$\mathbf{E}_s = \frac{\mu_0 c^2}{2\Delta^2} \left\{ -ik\mathbf{p}_P + ik\hat{\mathbf{R}} \times \left[ \mathbf{m}_P - \frac{k^2}{10}\mathbf{m}_P^{(1)} \right] - k^2 \left[ \mathbf{T}_P - \frac{k^2}{10}\mathbf{T}_P^{(1)} \right] + k^2 \left[ \mathbf{Q}^{(e)}.\hat{\mathbf{R}} \right]_P - \frac{k^2}{2}\hat{\mathbf{R}} \times \left[ \mathbf{Q}^{(m)}.\hat{\mathbf{R}} \right]_P \right. $$
$$\left. -\frac{ik^3}{3}\left[ \mathbf{Q}^{(T)}.\hat{\mathbf{R}} \right]_P - ik^3\left[ (\mathbf{O}^{(e)}.\hat{\mathbf{R}})\hat{\mathbf{R}} \right]_P - \frac{ik^3}{180}\left[ (\mathbf{O}^{(m)}.\hat{\mathbf{R}})\hat{\mathbf{R}} \right]_P \right\} \times \exp(-ikR) \quad (7)$$

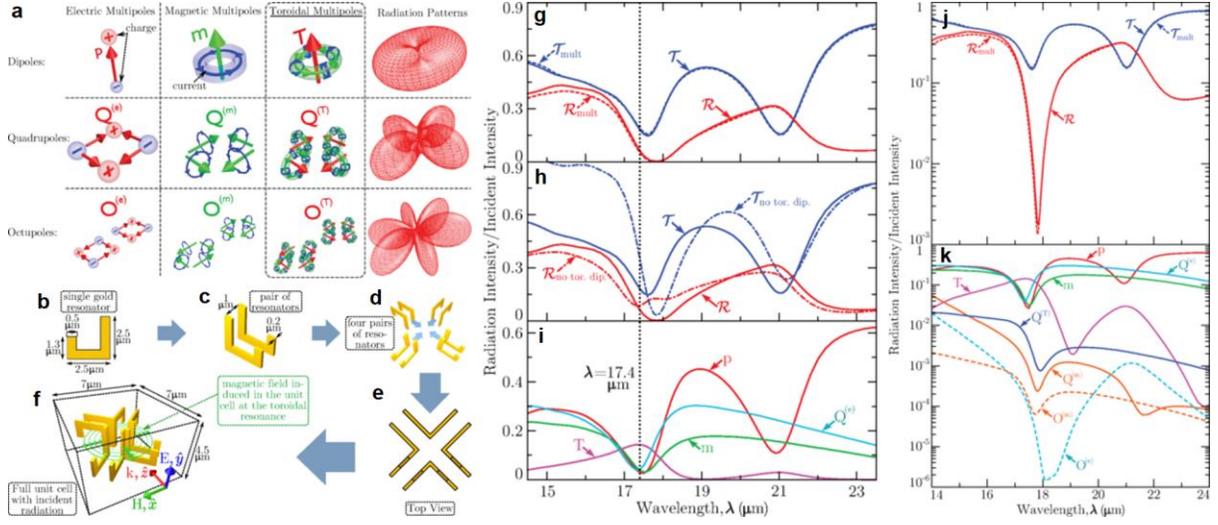

**Figure 14.** a) The families of dynamic multipoles. The three columns on the left show the charge-current distributions correlating with the electric (**p**), magnetic (**m**), and toroidal (**T**) dipoles, electric (**Q**$^{(e)}$), magnetic (**Q**$^{(m)}$), and toroidal (**Q**$^{(T)}$) quadrupoles, as well as the electric (**O**$^{(e)}$), magnetic (**O**$^{(m)}$), and toroidal (**O**$^{(T)}$) octupoles. The column on the right shows the patterns of radiation (i.e., intensity as a function of direction) emitted by the various harmonically oscillating multipoles. b) An L-shaped resonator out of gold. c) The L-shaped resonators are put in pairs with 1-μm gap distance between them. d, e) Four pairs of resonators are combined to make a single large resonator. f) The full unit cell which consists of the large gold resonator at the center embedded into the SU-8 polymer. g) Transmission and reflection spectra for the metamolecule. h) The comparison of the numerically and theoretically computed transmission and reflection to the response obtained from multipoles without the contribution of toroidal dipole. i) The intensity of the radiation scattered from each four distinct multipoles in arrays. j) The response of the toroidal metamaterial. Extended series of multipoles that make up the response of the metamaterial same as (g). k) The intensity reflected by arrays of multipoles induced in the metamaterial by incident radiation.[136] Copyright 2014, American Physical Society.

This equation gives rise to qualitative computation of the electric field emitted by a surface composed of toroidal scatterers. The contributed terms to the emitted field are electric (**p**), magnetic (**m**), and toroidal dipoles (**T**), electric ($\mathbf{Q}^{(e)}$), magnetic ($\mathbf{Q}^{(m)}$), and toroidal ($\mathbf{Q}^{(T)}$) quadrupoles, electric($\mathbf{O}^{(e)}$), magnetic($\mathbf{O}^{(m)}$), and toroidal ($\mathbf{O}^{(T)}$) octupoles. The mean-square radii of toroidal and magnetic dipoles are denoted by $\mathbf{T}^{(1)}$ and $\mathbf{m}^{(1)}$, respectively. It should be underline that the terms for multipolar modes above octupole (i.e. hexadecapole, etc.) can be neglected due to weak impact on the total emitted electric field.[136,139,142] The projected toroidal dipole ($\mathbf{T}_P$) from a planar scatterer located along the *xy*-plane is given by: $\mathbf{T}_P = T_x\hat{\mathbf{x}} + T_y\hat{\mathbf{y}}$ or in general: $\mathbf{T}_P = \mathbf{T} - (\mathbf{T}\cdot\hat{\mathbf{R}})\hat{\mathbf{R}}$. This allows to derive the far-field distributions for other isolated multipoles by using the expression for the radiation emitted by single multipole sources. Employing Eqn. 7, one can determine the radiation reflected and transmitted by planar arrays of scatterers under normal electromagnetic wave excitation.

To study the model provided for the toroidal metamaterials, Savinov and colleagues have tailored a multipixel unit cell consisting of 3D resonators encompassed by the SU-8 polymer, as shown in Figures 14b-14f, which includes the description for the geometrical sizes and orientation of each component.[136,238] The incident beam direction and the corresponding polarization is demonstrated in Figure 14f. The numerically (solid line) and theoretically (dashed line) calculated transmission and reflection profiles are depicted in Figures 14g and 14h, showing the formation of two resonant modes at 17.4 µm and 21 µm. To indicate the nature of the induced modes and validate the excitation of toroidal moments, one should carry out the multipole scattering analysis similar to the profile plotted in Figure 14i. It is obvious that at 21.0 µm, the metasurface response is dominated by the electric quadrupole and magnetic dipole scattering, and thus this resonance will be of no interest for the toroidal purpose. On the other hand, at 17.4 µm, the metamaterial response is clearly dominated by the toroidal dipole scattering which is more than three times larger than the contribution due to any other multipole. This figure also exhibits that the toroidal dipole excitation must play a key role in forming the metamaterial macroscopic response at the shorter-wavelength resonance (17.4 µm). This can be confirmed directly by ignoring the toroidal dipole moment in the multipole calculations of the transmission and reflection responses. Moreover, Savinov and colleagues have studied the intensity of reflected by multipoles as plotted in Figures 14j and 14k. Accordingly, the spectral response of the proposed metamaterial is accurately computed by multipole expansion up to order $k^4$.

## 5.2. Dynamic toroidal dipoles for refractive index sensing

Similar to previous studies, we first quickly review the recent studies in the field of the targeted resonant metasurfaces for RI sensing. As a pioneer work, Gupta *et al.*[248] have employed plasmonic toroidal metasurfaces for RI sensing with high sensitivity and FOM values. To this end, an antisymmetric multipixel high-*Q* planar toroidal metamolecule was developed to operate at THz frequencies (Figure 15a). The size of each pixel is specified in the picture and center-to-center gap size between proximal resonators was fixed to 9 µm. Figure 15b is a microscopic image from the fabricated sample. As discussed in the previous section, in the multipixel unit cells, a mismatch between the excited magnetic fields in neighbor resonators leads to excitation of toroidal moments and formation of head-to-tail magnetic field, as shown in Figure 15c. Clearly, under the normal beam excitation in the *x*-direction, toroidal dipolar mode ($T_x$) forms according to right hand rule. Focusing on the sensing properties of the proposed metasurface by Gupta and co-workers, by adding an analyte layer with the RI of *n*=1.66 (Figure 15d), they estimated the resonance shift due to this change. Figure 15e illustrates the excitation and shift of toroidal dipolar mode as a function of incident beam frequency for the absence and presence of analyte cover layer. The narrow lineshape of the toroidal dipole allows to determine the resonance shift due to the presence of high-index layer precisely. In addition, to study the influence of the toroidal mode in the sensing application, the researchers computed the *x*-component of the toroidal mode ($T_x$) for the presence and absence of analyte layer (Figure 15f). In the presence of analyte, a distinct red-shift in the position of toroidal mode is recorded due to red-shift in the toroidal coupling between proximal resonators.

To experimentally show the sensitivity of the toroidal system to the surrounding media perturbations, Gupta and teammates employed two different coating layers ($SiO_2$ and Germanium (Ge)) with high RI contrast, with the thickness of 250 nm. Figures 15g and 15h show the transmission spectra for the simulated and experimentally analyzed RI variations, respectively. In this regime, by increasing the RI of the medium and moving from Nitrogen ($N_2$) with the RI of *n*=1 to Ge with the index of *n*=4, the researchers monitored a continuous red-shift in the position of toroidal dipole. Here the corresponding sensitivity can be defined by plotting the resonance shift as a function of RI alterations, as demonstrated in Figure 15i ($\left| d\lambda/dn \right| = c/f_0^2 \left( df/dn \right)$). The results revealed the sensitivity of 6 GHz/RIU for the studied metadevice. However, by increasing the thickness of the analyte layer from 250 nm to thicker layers, one can obtain sensitivity up to 24 GHz/RIU. To provide a proof for this this claim, the sensitivity of the toroidal resonance for thicker analyte layers (4 µm) and various RIs have been performed and depicted in Figure 15j. Similar to earlier studies, by gradual increases in the RI of the media, a continuous red-shift is observed in the position of toroidal mode. Figure 15k demonstrates the frequency shift as a function of RI variations and the slope of the line clearly indicates the sensitivity of 27.3 GHz/RIU for the toroidal system. In addition, Gupta *et al*. have verified that increasing the thickness of analyte coverage up to 25 µm gives rise to the significant enhancement in the corresponding sensitivity of the metasystem up to 186 GHz/RIU ($10.3\times10^4$ nm/RIU). It should be noted that the obtained sensitivity in the recent work is highly promising and opens new doors to employ toroidal metasurfaces for practical biomolecule sensing purposes with high precision.

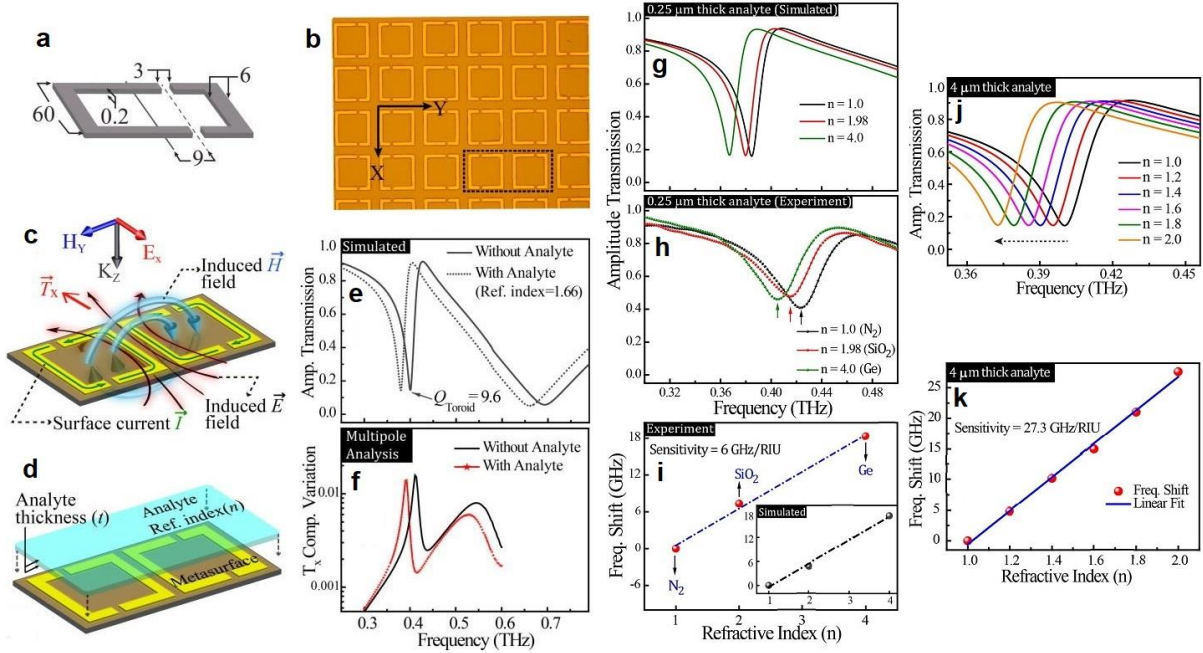

**Figure 15.** a) Dimensional features of the THz asymmetric pixel. b) Microscopic image of toroidal metasurface along the *xy*-plane. c) Schematic of toroidal dipole generated due to the circulating magnetic field produced by surface currents induced in the configuration. d) Unit cell of the toroidal metamolecule coated with the analyte layer on the top. e) Simulated frequency versus amplitude transmission spectra of the metasurface with and without the analyte layer. f) Computed obtained *x*-component of the toroidal dipole moment as a function of frequency with and without analyze cover. g) Simulated and h) experimentally measured amplitude transmission spectra showing the shift in toroidal resonance with different analytes. i) Dipolar resonance frequency shift versus RI plot for samples coated with $SiO_2$ and Ge. j) Simulated amplitude transmission spectra, and correspondingly derived k) frequency shift against RI for the 4 µm thick analyte layer, with an increasing index, coated on the toroidal metasurface.[248] Copyright 2017, American Institute of Physics.

## 5.3. Dynamic toroidal dipole-resonant metasensors for biological sensing

As discussed in prior sections and subsections, unconventional characteristics of toroidal multipoles have stimulated researchers to study these unique resonant phenomena by using both 3D and planar resonators to develop advanced nanophotonic metadevices. In the latest subsection, we reviewed very recent updates on the use of toroidal dipoles for RI sensing purposes. However, in terms of practical applications for immunosensing and real-time pharmacology applications, new original studies have shown that how toroidal metasurfaces enable to recognize ultra-low weight biomolecules and proteins using toroidal metamaterials technology.[132,133] Ahmadivand and colleagues have experimentally verified that plasmonic toroidal metasurfaces can be tailored to detect Zika virus (ZIKV) protein efficiently with the molecular weight of 13 kDa. To this end, a multimetallic multipixel metamolecule has been utilized to support pronounced toroidal dipole along the THz spectrum. Promisingly, THz spectroscopy enables for noninvasive, noncontact, nondestructive, and label-free biomarker detection in biomedical and clinical applications. Figures 16a and 16b illustrate schematic and top-view images for the developed metamolecule and the corresponding optimal geometries are reported in the figure's caption. The SEM images for the fabricated unit cell is also demonstrated in Figures 16c and 16d. The central bar was set to Titanium (Ti) and the peripheral resonators were considered to be Iron (Fe). Figures 16e and 16f illustrate the spectral response of the metamolecule under intense normal beam radiation, experimentally and numerically, respectively. These profiles verify the excitation of pronounced toroidal dipole around 203 GHz with experimental *Q*-factor of 18. It is worthy to note that another resonant dip is also induced around 230 GHz correlating with the magnetic dipolar moment arising from the peripheral curved resonators that is not the concept of interest in the sensing mechanism. Ahmadivand and teammates have shown that the exquisite electromagnetic response of the studied THz metasurface can be employed to tailor highly sensitive and accurate plasmonic sensors. The immunosensing specimens have been prepared using following methods: (1) with antibody, (2) with antibody and bovine serum albumin (BSA), and (3) with antibody, bovine serum albumin (BSA), and variant concentration (ranging between 1 pg/mL to $10^4$ pg/mL) of immobilized ZIKV envelope protein (The full description is provided in the Methods section in Ref. 132). The SEM images for this process and the presence of immobilized ZIKV antibody on a metamolecule and a chip covered with antibody-attached ZIKV envelope protein are illustrated in Figures 16g and 16k, respectively. The spectral results and the behavior of toroidal dipole to the presence of antibody, and biomarker proteins are demonstrated in Figure 17. In this panel, the shift in the toroidal mode position is demonstrated as a function of frequency for various concentrations of ZIKV envelope proteins (Figure 17a). Once can clearly see huge red-shift in the resonance position (from 203 GHz to 198 GHz) by increasing the concentration of targeted bioproteins. However, further increases in the concentration of ZIKV proteins to 50 pg/mL and 100 pg/mL, shift the toroidal resonance to 194 and 188 GHz, respectively. Such a large shift in the resonance frequency shows the sensitivity of the toroid dip to the concentration of the infection protein. Ultimately, experimental assays proved that extra increases in the concentration of biomarkers envelope protein to 500 pg/mL leads to a shift of the position of the pronounced toroidal resonant dip to 187 GHz. In addition, by increasing the concentration of target protein to $10^3$ pg/mL and $10^4$ pg/mL, a drastic decay has been monitored in the quality of the toroidal mode in both cases. Figure 17b exhibits the toroidal frequency shifts as a function of envelope protein concentration (pg/mL), showing how the metasensor is sensitive to the protein concentrations. This graph has been employed to calculate the corresponding sensing parameters. Thus the LOD of plasmonic sensor has been defined by LOD = 3(*SD*)/*S*, where "*SD*" is the standard deviation of the frequency shift and "*S*" is the slope of the fitting line (shown by the dashed line in Figure 17b), and the LOD was quantified as ∼24 pg/mL. By defining the slope of the toroidal position shift as a function of envelope protein concentration, once can assessed the sensitivity of the structure as 6.47 GHz/log(pg/mL).

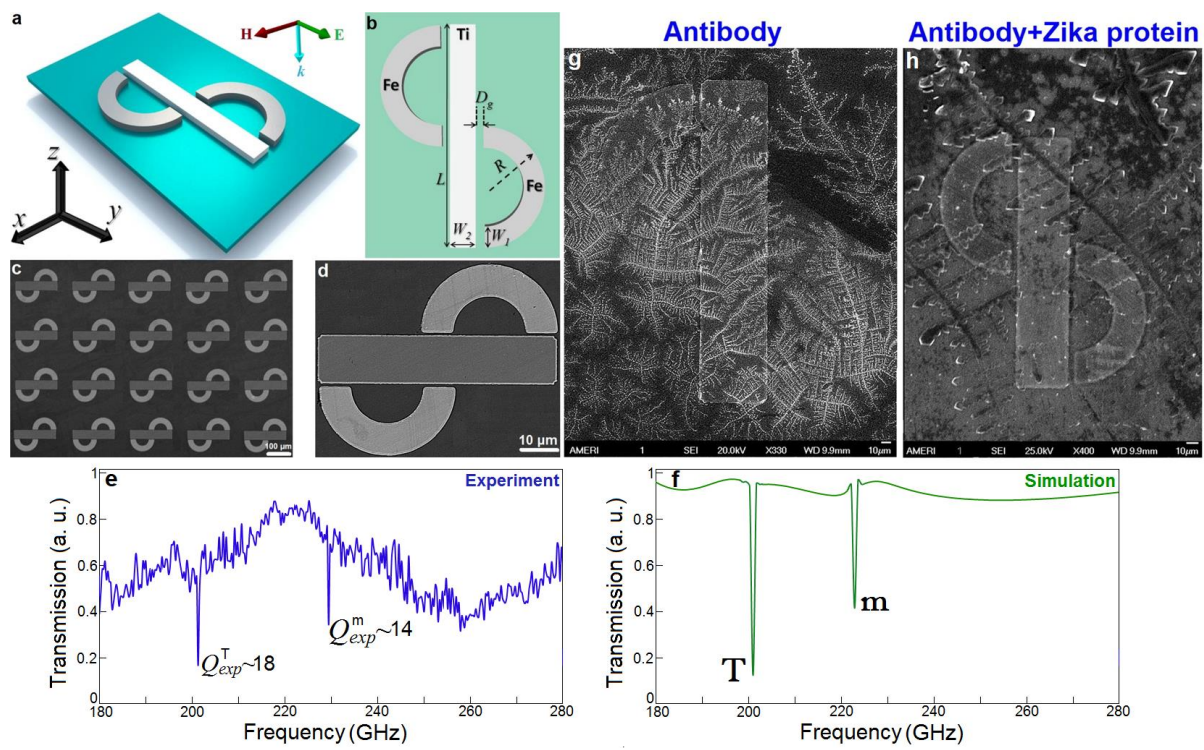

**Figure 16.** a) Artistic schematic of multimetallic plasmonic resonators assembly as a unit cell on a silicon host. b) Top-view image of the toroidal unit cell with an introduction to geometrical components. c) SEM image of fabricated plasmonic structures in arrays for the unit cells with the gap spots between surrounding and central resonators of $D_g$= 3 μm with $L$= 240 μm, $R$= 50 μm, $W1$= 30 μm, and $W2$= 40 μm. d) The magnified SEM image for each unit cell with $D_g$= 3 μm. g, h) SEM images of the plasmonic toroidal resonator covered with antibody and ZIKV envelope proteins attached to the antibody, respectively.[132] Copyright 2017, American Chemical Society.

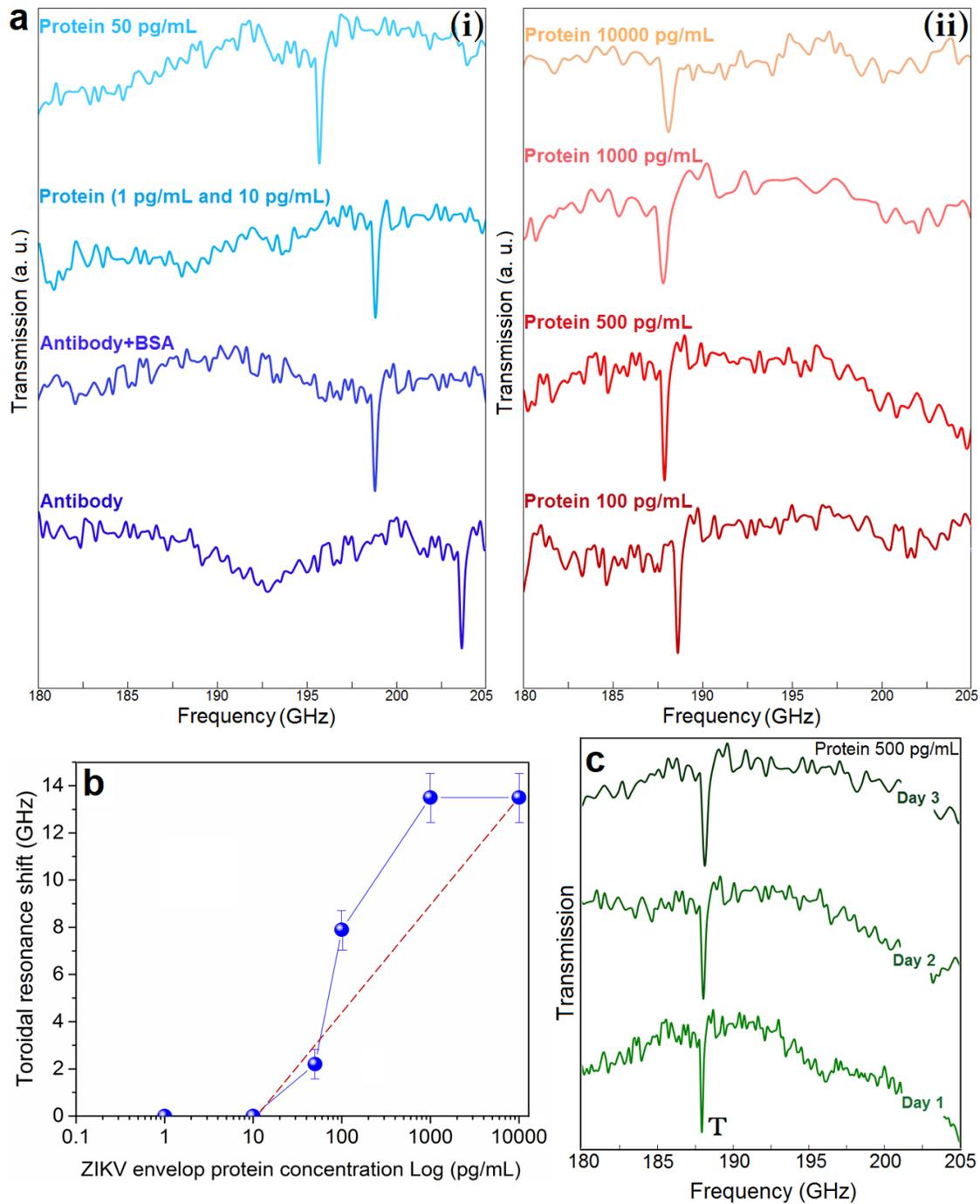

**Figure 17.** a) Transmission spectra for the toroidal resonant mode behavior for presence of different concentration of ZIKV envelope protein from (i) antibody to 50 pg/mL and (ii) 100 pg/mL to $10^4$ pg/mL. b) Toroidal resonance frequency shifts due to conjugated ZIKV protein concentration (solid) and fitting line (dotted). c) Transmission spectra for a THz plasmonic chip characterized for 3 days to define the repeatability of a sample.[132] Copyright 2017, American Chemical Society.

Among all sensing parameters, the repeatability of biosensors plays a critical role in developing efficient and cost-effective devices. Ahmadivand *et al*. have shown that the developed plasmonic toroidal samples can be employed for three days, which holding same spectral response (Figure 17c). This figure indicates the measured transmission spectra for three consecutive days with the presence of the envelope protein with the concentration of 500 pg/mL. It should be noted that after this period of time, the toroidal dip became broader and dramatically damped. Here the researchers successfully verified the detection of proteins with the ultra-low molecular weight of ≈13 kDa, while the recent achievements show detection of bio-objects with the weight of over 70 kDa using plasmonic biosensors for pM concentrations.

These results reveal the fact that among all of the referenced and comprehensively reviewed achievements, toroidal metasensors show much superior performances and faster responses in terms of the concentration and low-molecular weight of bio-agents. However, this is not the end of story, and by utilizing toroidal plasmonic metasurfaces, it has successfully been verified and shown that plasmonic toroidal metasensors can be optimized in terms of binding strength of biomolecules with unit cells, as well as the sensitivity, and repeatability of developed sensors. The other very recent proof of concept for the use of toroidal metamaterials for biological sensing was proposed and studied by Ahmadivand *et al.*[133] have shown that integration of Au NPs with toroidal metamolecules can enhance the sensing performance of plasmonic metasensors. Again by focusing on the THz band of the spectrum, as a promising bandwidth for biosensing purpose, the researchers in this article employed a multipixel symmetric unit cell (schematic in Figure 18a).

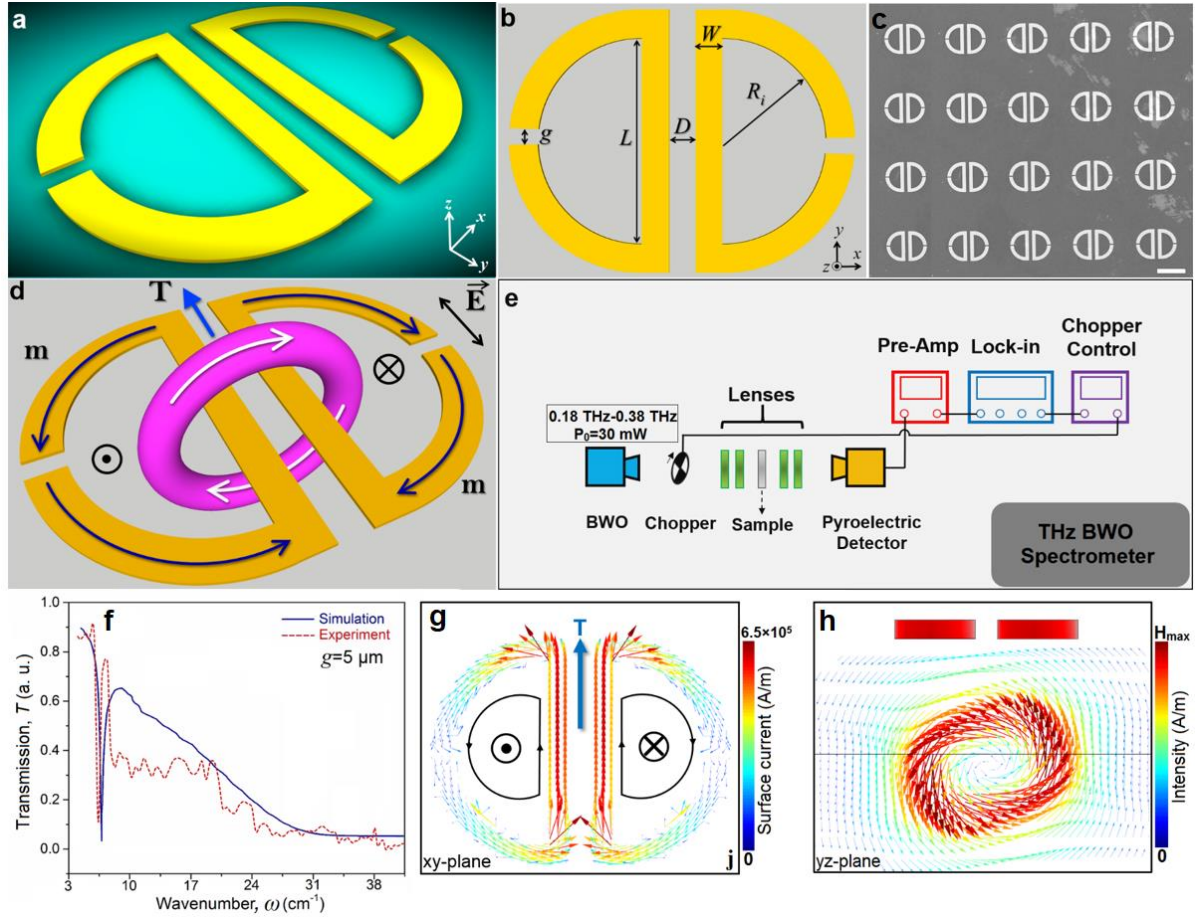

**Figure 18.** a) Schematic of the toroidal unit cell. b) Description to the geometrical parameters of the unit cell. c) SEM image of the fabricated metasurface. d) Schematics of the formation of head-to-tail arrangement correlating with the toroidal field between the adjacent resonators e) Schematic of the BWO setup, utilized to characterize the spectral response of the metamolecule. f) Normalized transmission amplitude for the toroidal metamaterial with following geometries: $R_i/W/L/D/D_g$ = 60/15/105/15/5 μm. g) Surface current map for the induced current flux across the structure and formation of toroidal mode. h) A cross-sectional yz-plane of the resonators, showing the head-to-tail magnetic moments forming the toroidal moment in a vectorial map.

The introduction to the associating geometrical components and SEM image are depicted in Figures 18b and 18c, respectively. Under normalized *y*-direction incidence, the formation of a head-to-tail arrangement correlating with the toroidal mode is expectable due to formation of opposite magnetic (**m**) fields in proximal resonators. The backward wave oscillator (BWO) setup to analyze the spectral behavior of the metasurface is demonstrated in Figure 18e. The numerical and experimental spectral results for the transmission spectra are plotted in Figure 18f, displaying formation of a distinct toroidal dipole peak around $\omega \sim 6.67$ cm$^{-1}$. The vectorial surface current map across the plasmonic unit cell at the toroidal dipole frequency validates the required mismatch between proximal magnetic fields (Figure 18g). The formation of pronounced toroidal dipole can be strongly verified by looking at the cross-sectional vectorial magnetic-field panel for the toroidal unit cell, as shown in Figure 18h, in which the formation of strong head-to-tail arrangement is clear. Obeying the same theory, as explained in the previous studies, the physics behind the excitation of toroidal dipole can be found in Ref. 133.

Next, we take a look at the practical application of high-*Q* toroidal dipole, studied by Ahmadivand and colleagues. Similar to the previous toroidal metasensor, here also ZIKV protein with ultra-low molecular weight has been employed as a target protein for sensing and showing the capabilities of the metasurface. In this context Au NPs have been used to integrate with the toroidal unit cell to enhance the performance of the tailored sensor. Figure 19a demonstrates a flowchart for the preparation of Au NPs and conjugation of antibody to the particles with the diameter of 40 nm to deposit on the plasmonic metasurface (Figure 19b). There NPs were used to trap the proteins (see Figure 19c and 19d) and change the RI of the medium to facilitate substantial shift in the position of toroidal dipole. For easiness, the authors used fixed concentration of NPs (10 μg/mL or ~77 pM), while the injected envelope protein concentration was varied between 1 fg/mL and 1 μg/mL. Figure 19d illustrates the SEM image of the toroidal unit cell with the presence of ZIKV envelope proteins, captured by functionalized NPs. To show the binding quality between antibody and proteins to the Au NPs, the corresponding TEM images have been captured and illustrated in Figure 19e-19g to show the uniform and strong binding of the bio-molecules to the NPs, which enhances the sensitivity of the metasensor.

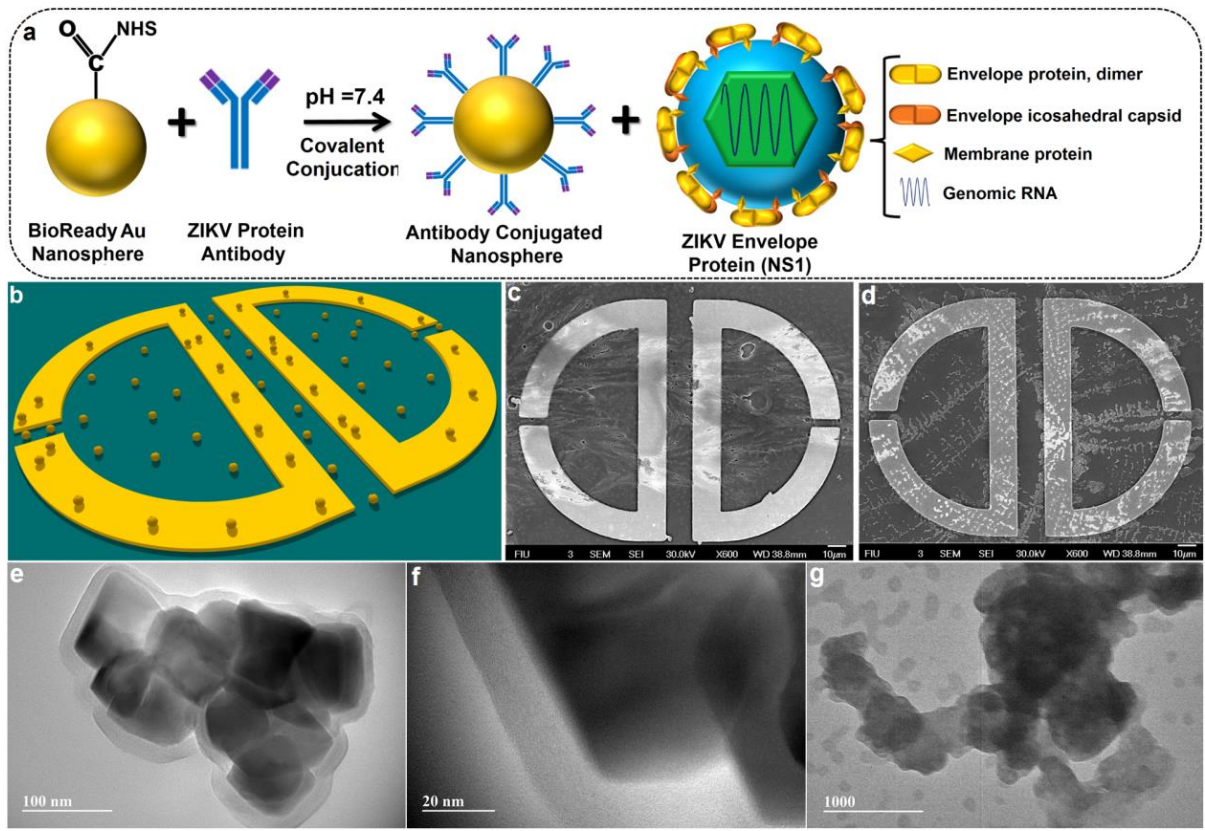

**Figure 19.** a) Schematic roadmap of functionalizing Au NP conjugation with the ZIKV antibody and the corresponding proteins with the explanation for different parts. b) Schematic representation of Au nanoparticles-integrated toroidal metamolecule. c, d) The SEM images of plasmonic metamolecule in the presence of Au NPs with antibody and ZIKV envelope proteins, respectively. e, f) TEM images of functionalized Au NPs binding with antibody in two different scales. (c) The TEM image of ZIKV envelope proteins captured by antibody-conjugated Au NPs.

The experimentally obtained THz transmission spectra with distinct toroidal responses for different concentrations of envelope proteins are presented in Figure 20a. For the absence of colloidal Au NPs and biological agents and in the air condition with the index of $n=1$, the magnetic toroidal dipole appears at 6.67 cm$^{-1}$. By introducing the antibody-conjugated Au NPs to the metamolecule chips, the toroidal mode red-shifts to 6.53 cm$^{-1}$. Conversely, in the absence of NPs and direct binding of antibody to the metamolecule, the resonant mode slightly shifts to 6.6 cm$^{-1}$. As the researchers reported, addition of targeted proteins with increasing concentration from 1 fg/mL to 100 pg/mL did not cause significant shift in the toroidal mode position and it remained at the original position of the presence of respective antibody in both assays. The substantial shift in the position of toroidal resonance was observed for 1 ng/mL concentration of biomarker proteins for the presence of Au NPs-conjugated system (moves to 6.33 cm$^{-1}$), although the toroidal lineshape did not shift for the bare metamolecule with the presence of the same amount of envelope proteins. It is shown that further increases in the concentration of biomarker proteins leads to much more red-shift in the position of toroidal dipole, while for the presence of 100 ng/mL of proteins, the toroidal mode appears at 6.13 cm$^{-1}$. On the other hand, in continue, the unit cell without any biological object starts to react to the presence of envelope proteins at the concentration of 10 ng/mL (red-shifts to 6.47 cm$^{-1}$). Besides, for the concentration of 100 ng/mL of targeted envelope proteins, the toroidal dipole red-shifts slightly to 6.4 cm$^{-1}$. The noteworthy point here the position of toroidal moment, which reflects minor shift for the concentrations of proteins more than ≥ 1 µg/mL. The toroidal dipole position displacement ($\Delta\omega$) as a function of envelope protein concentration is plotted in a semi-logarithmic diagram in Figure 20b for the presence and absence of Au NPs. This profile exhibits how the presence of Au NPs improves the sensitivity of the toroidal metasensor. Clearly, a large red-shift was obtained for Au NPs-enhanced system around $\Delta\omega$~0.35 cm$^{-1}$ or $\Delta f$ ~10 GHz, (Figure 20c) for the biomarker concentration of 100 ng/mL (~77 pM) with the coefficient of determination $R^2$= 0.982. For the absence of NPs, the biggest shift was around $\Delta\omega$~0.14 cm$^{-1}$ or $\Delta f$ ~4 GHz, (Figure 20c), with the corresponding determination coefficient of $R^2$= 0.88. Ultimately, Ahmadivand *et al*. have shown that how the presence of Au NPs attached to the toroidal system optimizes the repeatability and stability of the metasensor. Figures 20d and 20e quantitatively and qualitatively compare the toroidal resonance shift and narrowness as a function of time at 100 ng/mL of envelope proteins. For the Au NPs-integrated metamolecule, the resonance shift remains unchanged for almost 60 hours, while a continuous blue-shift in the position of toroidal mode is monitored. This blue-shift and decay in the toroidal mode quality are much more drastic in the absence of Au NPs (Figure 20b). The same behavior has been detected for the *Q*-factor, in which the *Q*-factor of the unit cells monotonically weakened. This deterioration also included a substantial blue-shift in the position of dipolar resonant mode to the higher energies. Lastly, the sensing parameters for the metadevice have been quantified accurately. Accordingly, the LOD and sensitivity of the metasensors have been reported as ~560 pg/mL and 5.81 GHz/log(pg/mL) for the Au NP-integrated metasensor and 12 ng/mL and 2.25 GHz/log(pg/mL) for the bare metasurface. This validates the key role of Au NPs in enhancing the sensitivity of the toroidal plasmonic metasensor.

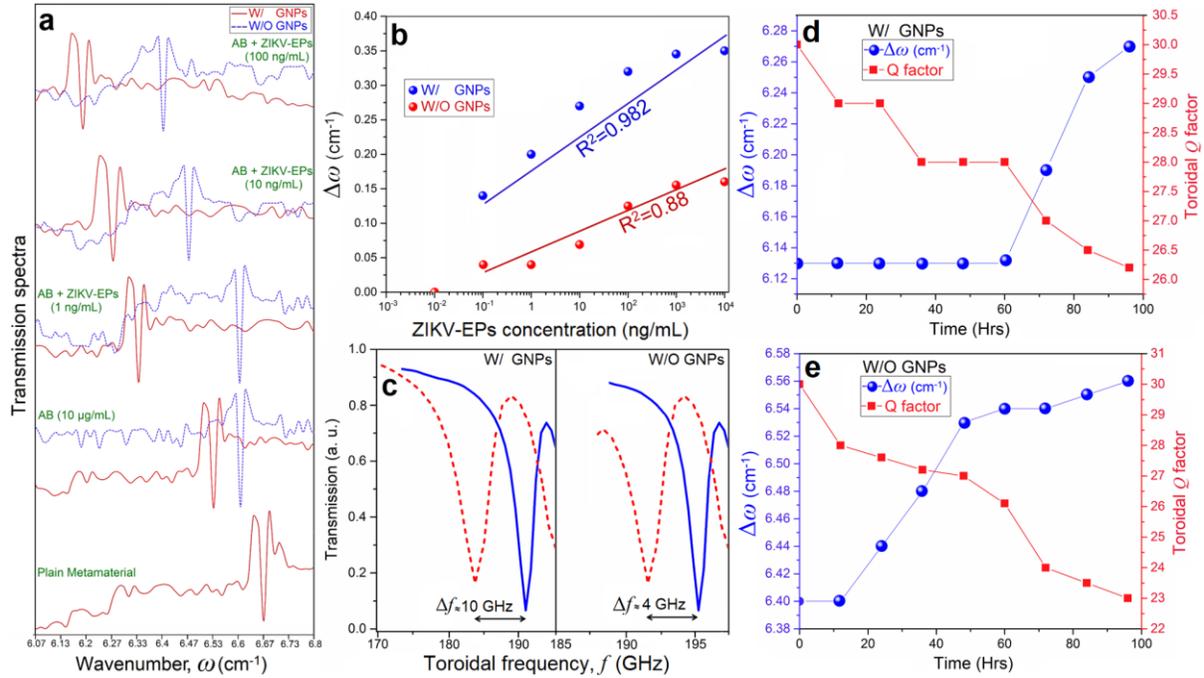

**Figure 20.** a) The transmission amplitude spectra for the fabricated metasurfaces in both W/ and W/O Au NPs regimes in the presence of antibody and proteins with different concentrations. b) The toroidal resonance shift as a function of envelope proteins concentration W/ and W/O Au NPs with the corresponding determination coefficient. c) The magnified transmission spectra as a function of frequency, showing the maximum shift of the toroidal moment in the presence and absence of Au NPs attached to the system. d, e) The toroidal resonance shift ($\Delta\omega$) and $Q$-factor as a function of time (in hours) for the presence and absence of Au NPs, respectively.

## 6. Conclusion and summary

As a conclusion and summary, we reviewed the recent achievements in the field of label-free biosensing using plasmonic NP aggregates and metamaterials. In this context, we qualitatively and quantitatively reviewed and evaluated the performance of existing classical plasmonic nano and microscale biosensors and recently proposed ones based on ghost resonance technology. Our investigations over the recent advances on the plasmonic sensing tools revealed the undeniable role of newly introduced toroidal resonances on both spectral and sensitivity of plasmonic detection tools for developing practical and modern laboratory sensors. To show this advantage, a comprehensive report was given on the operation mechanism and sensing capabilities of some of colloidal NPs-based and well-engineered classical plasmonic label-free sensing platforms. It is shown that classical plasmonic NPs supporting DLSPR modes have been employed for various protein and biomolecule detection approaches with limited sensitivity and at high concentration of biological targets. On the other hand, we have discussed the benefits of using well-engineered Fano-resonant and perfect absorber metasurfaces for real-time monitoring of lipid–protein systems in aqueous environments. However, despite bearing promise for advance biosensing purposes, these subwavelength structures are quite far from the real-life needs. Relatively, the shortcoming of these systems have also described, especially reasonable sensitivity, poor LOD, and limited repeatability. Considering the fact that early detection of specific viruses and associating agents plays fundamental and critical role in anticipation and prevention of several disease. Ultimately, we perceived that novel type of metasurfaces with the ability to support a third family of multipoles (toroidal modes) have been reported as a promising technique to develop immunosensing devices with high precision and excellent LOD. It has been proved that toroidal-resonant metasensors enable detection of ultra-low molecular weight bio-agents at pmol concentrations with 100-fold sensitivity enhancement in comparison to analogous traditional devices. The theoretical studies and experimental assays confirmed that such a promising performance relies on the ultrahigh sensitivity of toroidal mode on the environmental RI value perturbations. We envision that this review paves new paths for the ones who are trying to develop new and advanced immunosensing tools based on toroidal plasmonic metasurfaces, as big rivals for conventional multiresonant plasmonic sensors and transducers.[198,249]

## Additional Information

**Competing financial interest**: The authors declare no competing financial interest.